\documentclass[showpacs,nofootinbib,showkeys,eqsecnum,prd,aps,preprint]{revtex4-1}
\usepackage{amsfonts,amsmath,amssymb,bm,graphicx,fontenc}
\usepackage[english]{babel}

\begin{document}

\title{{\large Particle scattering and vacuum instability by exponential
steps}}
\author{S.~P.~Gavrilov${}^{a,c}$}
\email{gavrilovsergeyp@yahoo.com, gavrilovsp@herzen.spb.ru}
\author{D.~M.~Gitman${}^{a,b,d}$}
\email{gitman@if.usp.br}
\author{A.~A.~Shishmarev$^{a,d}$}
\email{a.a.shishmarev@mail.ru}
\date{\today}

\begin{abstract}
Particle scattering and vacuum instability in a constant inhomogeneous
electric field of particular peak configuration that consists of two
(exponentially increasing and exponentially decreasing) independent parts
are studied. It presents a new kind of external field where exact solutions
of the Dirac and Klein-Gordon equations can be found. We obtain and analyze
in- and out-solutions of the Dirac and Klein-Gordon equations in this
configuration. By their help we calculate probabilities of particle
scattering and characteristics of the vacuum instability. In particular, we
consider in details three configurations: a smooth peak, a sharp peak, and a
strongly asymmetric peak configuration. We find asymptotic expressions for
total mean numbers of created particles and for vacuum-to-vacuum transition
probability. We discuss a new regularization of the Klein step by the sharp
peak and compare this regularization with another one given by the Sauter
potential.
\end{abstract}


\affiliation{${}^{a}$Department of Physics, Tomsk State University, 634050 Tomsk, Russia;\\
${}^{b}$P.N. Lebedev Physical Institute, 53 Leninskiy prospect, 119991 Moscow,
Russia;\\
${}^{c}$Department of General and Experimental Physics, Herzen State
Pedagogical University of Russia, Moyka embankment 48, 191186
St.~Petersburg, Russia\\
${}^{d}$Institute of Physics, University of S\~{a}o Paulo, Rua do Mat\~{a}o, 1371, CEP 05508-090, S\~{a}o Paulo, SP, Brazil}

\maketitle

\section{Introduction}

Particle creation from a vacuum by strong electromagnetic and gravitational
fields is a well-known quantum effect \cite{Sch}, which has a number of
important applications in laser physics, heavy ion collisions, astrophysics,
and condensed matter processes (see Refs.~\cite{Dun09,RufVSh10,GelTan15} for
a review). Depending on the strong field structure different approaches have
been proposed for nonperturbative calculating of the effect. In these
approaches the strong fields are considered as external ones (external
classical backgrounds). Initially, the effect of particle creation was
studied for time-dependent external electric fields that are switched on and
off at the initial and final time instants, respectively. We call such
external fields $t$-electric potential steps. Initially, scattering,
particle creation, and particle annihilation by $t$-electric potential steps
have been considered in the framework of the relativistic quantum mechanics,
see for example Refs. \cite{Nikishov0,Nikishov1}. At present it is well
understood that only an adequate quantum field theory (QFT) with a
corresponding external background may consistently describe this effect and
possible accompanying processes. In the framework of such a theory, particle
creation is related to a violation with time of a vacuum stability. In
quantum electrodynamics (QED), backgrounds that may violate the vacuum
stability are electriclike electromagnetic fields. A general nonperturbative
formulation of QED with $t$-electric potential steps was developed in Refs.
\cite{Git77,FradGit,FGS}. The corresponding technique uses essentially
special sets of exact solutions of the Dirac equation with the corresponding
external backgrounds. The cases when such solutions can be found explicitly
(analytically) are called exactly solvable cases. At the current moment, all
known exactly solvable cases for $t$-electric potential steps are studied in
detail, see Ref. \cite{t-case} for a review.

However, there exist many physically interesting situations where external
backgrounds are formally presented by time independent fields (which is
obviously some kind of idealization). For example, one can mention
time-independent nonuniform electric fields that are concentrated in
restricted spatial areas. Such fields represent a kind of spatial or, as we
call them, conditionally, $x$-electric potential steps for charged
particles. The $x$-electric potential steps can also create particles from
the vacuum; the Klein paradox is closely related to this process \cite%
{Klein,Sauter1,Sauter2}. Approaches for treating quantum effects in the $t$%
-electric potential steps are not applicable to $x$-electric potential
steps. Some heuristic calculations of particle creation by $x$-electric
potential steps in the framework of the relativistic quantum mechanics with
a qualitative discussion from the point of view of QFT were first presented
by Nikishov in Refs. \cite{Nikishov1,Nikishov2}. In the recent article \cite%
{x-case}, quantizing the Dirac and the Klein-Gordon (scalar) fields in the
presence of $x$-electric potential steps, Gavrilov and Gitman presented a
consistent nonperturbative formulation of QED with $x$-electric potential
steps. Similar to $t$-electric potential step case, special sets of exact
solutions of the Dirac equation with the corresponding external field are
used to form a base of this formulation. By the help of this approach
particle creation in the Sauter field $E(x)=E\cosh^{-2}\left( x/L_{\mathrm{S}%
}\right) $ and in the so-called $L$-constant electric field (a constant
electric field between two capacitor plates) were studied in Refs. \cite%
{x-case} and \cite{L-field}, respectively. These two cases are exactly
solvable for $x$-electric potential steps. In the present article, we
consider another new exactly solvable case of this kind, which is a constant
electric field of particular peak configuration. The corresponding field is
a combination of two exponential parts, one exponentially increasing and the
other one exponentially decreasing. Different choice of these two parts
allows one to imitate different realistic and physically interesting spatial
configuration of electric fields. Besides of this, a very sharp peak can be
considered as a field of a regularized Klein step. We compare this
regularization with one given by the Sauter potential in Ref. \cite{x-case}.

The article is organized as follows. In Sec. \ref{peak_field}, a general
form of the constant electric field of a peak configuration that consist of
two (exponentially increasing and exponentially decreasing) independent
parts is introduced. We obtain and analyze corresponding in- and
out-solutions of the Dirac and Klein-Gordon equations. By their help we
introduce initial and final sets of creation and annihilation operators of
electrons and positrons and define initial and final vacua. In Sec. \ref%
{scattering} we discuss scattering and reflection of particles outside of
the Klein zone while possible processes in the Klein zone are studied in
Sec. \ref{Klein}. Characteristics of the vacuum instability in the Klein
zone are calculated by the help of in- and out-solutions using results of a
general theory \cite{x-case}. Here a particular case of a small-gradient
field is discussed as well. In Sec. \ref{exp_dec_field} we study a strongly
asymmetric peak configuration. In Sec. \ref{Sharp_peak} we consider a very
sharp peak. Mathematical details of study in the Klein zone are placed in
Appendix \ref{App.1}. In Appendix \ref{App.2} some necessary asymptotic
expansions of the confluent hypergeometric function are given.

Throughout the article, the Greek indices span the Minkowski space-time, $%
\mu =0,1,\dots ,D$. We use the system of units where $\hslash =c=1$ in which
the fine structure constant is $\alpha =e^{2}/c\hslash =e^{2}$.

%%%%%%%%%%%%%%%%%%%%%%%%%%%%%%%%%%%%%%%%%%%%%%%

\section{In- and out-solutions in exponential steps \label{peak_field}}

\subsection{Dirac equation}

We consider an external electromagnetic field, placed in $(d=D+1)$%
-dimensional Minkowski space, parametrized by the coordinates $%
X=(X^{\mu},\mu =0,1,\ldots,D)=(t,r)$, $X^{0}=t$, $r=(x,r_{\bot})$, $%
r_{\bot}=\left( X^{2},\ldots,X^{D}\right) $. The potentials of an external
electromagnetic field are chosen as $A^{\mu}(X)=\delta_{0}^{\mu}A_{\mu}(x)$,%
\begin{equation}
A^{\mu}(X)=(A^{0}=A_{0}\left( x\right) \,,\ \ A^{k}=0\,,\ \ k=1,...,D)\,,
\label{f.1}
\end{equation}
which corresponds to the zero magnetic field and the electric field of the
form%
\begin{equation}
\mathbf{E}\left( X\right) =\mathbf{E}\left( x\right) =\left( E_{x}\left(
x\right) ,0,...,0\right) ,\ \ E_{x}\left( x\right) =-\partial_{x}A_{0}\left(
x\right) =E\left( x\right) .  \label{f.2}
\end{equation}
The electric field (\ref{f.2}) is directed along the $x$-axis, inhomogeneous
and constant in time in general case. The backgrounds of this kind represent
a kind of spatial $x$-electric potential steps for charged particles. The
main properties common to any $x$-electric potential steps are%
\begin{equation}
A_{0}\left( x\right) \overset{x\rightarrow\pm\infty}{\longrightarrow}%
A_{0}\left( \pm\infty\right) ,\ \ E\left( x\right) \overset{\left\vert
x\right\vert \rightarrow\infty}{\longrightarrow}0\,,  \label{f.3}
\end{equation}
where $A_{0}\left( \pm\infty\right) $ are some constant quantities, and the
derivative of the potential $\partial_{x}A_{0}\left( x\right) $ does not
change its sign for any $x\in\mathbb{R}$. For definiteness, we suppose that%
\begin{equation}
\frac{\partial A_{0}\left( x\right) }{\partial x}\leq0\Longrightarrow
\left\{
\begin{array}{l}
E\left( x\right) =-\partial_{x}A_{0}\left( x\right) \geq0 \\
A_{0}\left( -\infty\right) >A_{0}\left( +\infty\right)%
\end{array}
\right. \,.  \label{f.4}
\end{equation}

\begin{figure}[h]
\includegraphics
{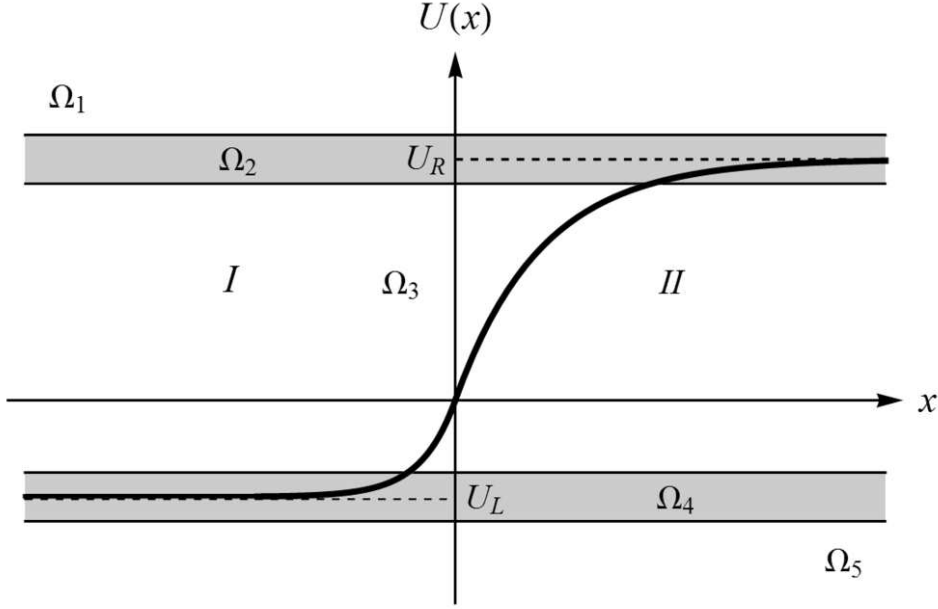}
\caption{Potential energy of electron $U(x)$ in peak electric field. For
this picture $k_{1}>k_{2}$ was chosen.}
\label{fig2}
\end{figure}

The basic Dirac particle is an electron, and the positron is its
antiparticle. The electric charge of the electron $q=-e$, $e>0$. The
potential energy of the electron in this field is $U(x)=-eA_{0}(x)$ (see
Fig. \ref{fig2}) and the magnitude of the corresponding $x$-potential step is%
\begin{equation}
\mathbb{U}=U_{\mathrm{R}}-U_{\mathrm{L}}>0,\ \ U_{\mathrm{R}%
}=-eA_{0}(+\infty ),\ \ U_{\mathrm{L}}=-eA_{0}(-\infty ).  \label{f.7}
\end{equation}%
One can distinguish two types of electric steps: noncritical and critical,
\begin{equation}
\mathbb{U}=\left\{
\begin{array}{l}
\mathbb{U<U}_{c}=2m\ , \\
\mathbb{U>U}_{c}\ ,%
\end{array}%
\right. \left.
\begin{array}{l}
\text{\textrm{noncritical steps}} \\
\text{\textrm{critical steps}}%
\end{array}%
\right. .  \label{f.7a}
\end{equation}%
In the case of noncritical steps, the vacuum is stable, see Ref. \cite%
{x-case}. We are interested in the critical steps, where there is
electron-positron pair production from vacuum.

\begin{figure}[h]
\centering
\includegraphics
{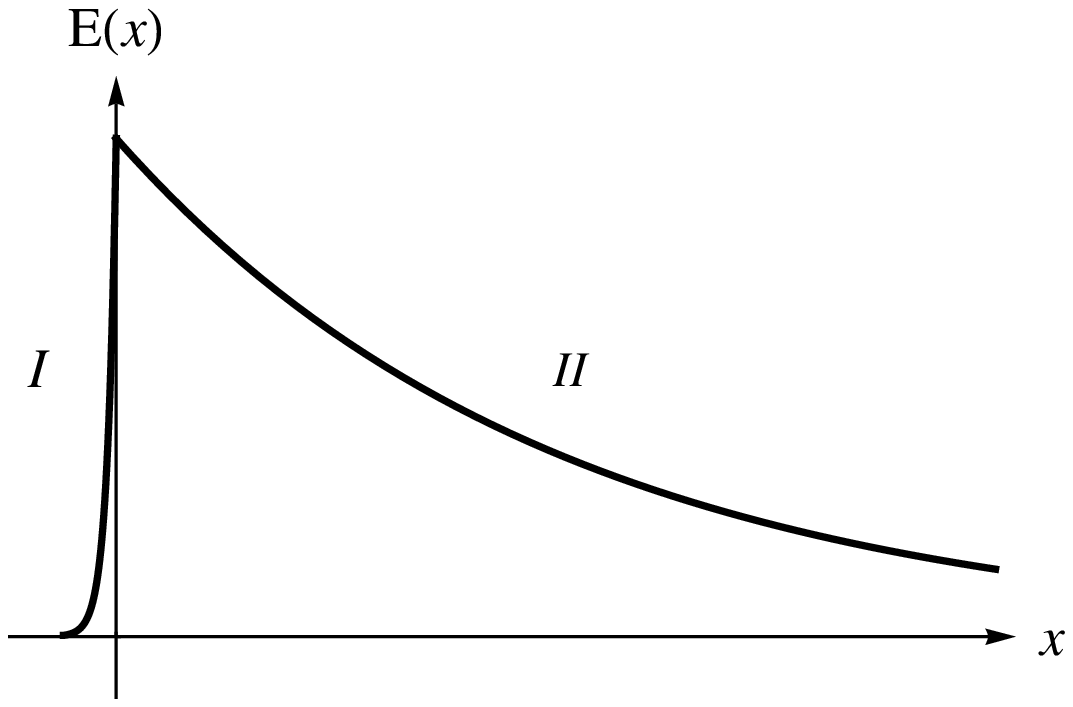}
\caption{Strongly asymmetric peak field configuration.}
\label{fig-as}
\end{figure}

System under consideration consists of a Dirac field $\psi(X)$ interacting
with the electric field of particular exponential configuration. This
electric field is composed of independent parts, wherein for each one the
Dirac equation is exactly solvable. The field in consideration grows
exponentially from the minus infinity $x=-\infty$, reaches its maximal
amplitude $E$ at $x=0$ and decreases exponentially to the infinity $%
x=+\infty $. Its maximum $E>0$ occurs at a very sharp point, say at $x=0$,
such that the limit%
\begin{equation}
\lim_{x\rightarrow-0}\frac{\partial E\left( x\right) }{\partial x}\neq
\lim_{x\rightarrow+0}\frac{\partial E\left( x\right) }{\partial x}\,,
\label{f.8}
\end{equation}
is not defined. The latter property implies that a peak at $x=0$ is present.
We label the exponentially increasing interval by $\mathrm{I}=\left(
-\infty,0\right] $ and the exponentially decreasing interval by $\mathrm{II}%
=\left( 0,+\infty\right) $. The field and its corresponding $x$-electric
potential step are
\begin{equation}
E\left( x\right) =E\left\{
\begin{array}{l}
e^{k_{\mathrm{1}}x}\,,\ \ x\in\mathrm{I} \\
e^{-k_{\mathrm{2}}x}\,,\ \ x\in\mathrm{II}%
\end{array}
\right. \,,\ \ U(x)=eE\left\{
\begin{array}{l}
k_{\mathrm{1}}^{-1}\left( e^{k_{\mathrm{1}}x}-1\right) ,\ \ x\in\mathrm{I}
\\
k_{\mathrm{2}}^{-1}\left( -e^{-k_{\mathrm{2}}x}+1\right) \,,\ \ x\in \mathrm{%
II}%
\end{array}
\right. \,,\   \label{f.9}
\end{equation}
where $E>0$, and$\ k_{\mathrm{1}},k_{\mathrm{2}}>0$; see Fig. \ref{fig-as}.
The potential energies of electron at $x=-\infty$ and $x=+\infty$ for this
particular configuration are%
\begin{equation}
U_{\mathrm{L}}=-\frac{eE}{k_{\mathrm{1}}},\ \ U_{\mathrm{R}}=\frac {eE}{k_{%
\mathrm{2}}}.  \label{f.9a}
\end{equation}

%figure 1
%%%%%%%%%%%%%%%%%%%%%%%%%
%%%%%%%%%%%%%%%%%%%%%%%%%

%figure 2
%%%%%%%%%%%%%%%%%%%%%%%%%
%%%%%%%%%%%%%%%%%%%%%%%%%

It should be noted that, for example, the strongly asymmetric peak
configuration, given by the potential%
\begin{equation}
A_{0}^{\mathrm{as}}\left( x\right) =E\left\{
\begin{array}{l}
0\,,\ \ \ \ \ \ \ \ \ \ \ \ \ \ \ \ \ \ \ \ \ \ x\in\mathrm{I} \\
k_{\mathrm{2}}^{-1}\left( e^{-k_{\mathrm{2}}x}-1\right) \,,\ \ x\in \mathrm{%
II}%
\end{array}
\right. \,,  \label{f.10}
\end{equation}
can be considered as a particular case of the step, that is, $k_{\mathrm{1}}$
is sufficiently large for this case, $k_{\mathrm{1}}\rightarrow\infty$.

The Dirac equation for the system under consideration has the following form:%
\begin{equation}
i\partial_{0}\psi(X)=\hat{H}\psi(X),\ \ \hat{H}=\gamma^{0}\left( -i\gamma
^{j}\partial_{j}+m\right) +U(x),\ \ j=1,\ldots D,  \label{de.1}
\end{equation}
where the Dirac field $\psi(X)$ is a $2^{[d/2]}$-component spinor (where $%
[d/2]$ is integer part of $d/2$) in $d$ dimensions, $\hat{H}$ is the
one-particle Hamiltonian, $\gamma^{\mu}$ are $2^{[d/2]}\times2^{[d/2]}$
gamma-matrices in $d$ dimensions (see, for example, Ref. \cite{BrauerWeyl}):%
\begin{equation}
\left[ \gamma^{\mu},\gamma^{\nu}\right] _{+}=2\eta^{\mu\nu},\ \ \eta^{\mu
\nu}=\text{diag}(\underset{d}{\underbrace{1,-1,\ldots,-1}}),\ \ \mu
,\nu=0,1,\ldots,D.  \label{de.2}
\end{equation}
Due to the configuration of the field (\ref{f.9}), the structure of Dirac
spinor $\psi(X)$ in directions $X^{0}$ and $X^{2},\ldots X^{D}$ is a simple
plane wave, so we consider stationary solutions of the Dirac equation (\ref%
{de.1}) having the following form
\begin{align}
& \psi_{n}(X)=\exp\left[ -ip_{0}t+i\mathbf{p}_{\bot}\mathbf{r}_{\bot }\right]
\psi_{n}(x),\ \ n=\left( p_{0},\mathbf{p}_{\bot},\sigma\right) ,  \notag \\
& \psi_{n}(x)=\left\{ \gamma^{0}\left[ p_{0}-U(x)\right] +i\gamma
^{1}\partial_{x}-\gamma^{\bot}\mathbf{p}_{\bot}+m\right\} \phi_{n}(x),
\notag \\
& \mathbf{p}_{\bot}=\left( p^{2},\ldots,p^{D}\right) ,\ \ \gamma^{\bot
}=\left( \gamma^{2},\ldots,\gamma^{D}\right) ,  \label{de.3}
\end{align}
where $\psi_{n}(x)$ and $\phi_{n}(x)$ are spinors that depend on $x$ alone.
These spinors are stationary states with the given total energy $p_{0}$ and
transversal momentum $\mathbf{p}_{\bot}$ (the index $\perp$ stands for the
spatial components perpendicular to the electric field). Spin variables are
separated by substitution%
\begin{equation}
\phi_{n}(x)=\phi_{n}^{\left( \chi\right) }(x)=\varphi_{n}(x)v_{\chi,\sigma
},\   \label{de.5}
\end{equation}
where $v_{\chi,\sigma}$ is the set of constant orthonormalized spinors with $%
\chi=\pm1$, $\sigma=\left( \sigma_{1},\ldots,\sigma_{\lbrack d/2]-1}\right) $%
, $\sigma_{s}=\pm1$, satisfying the following relations%
\begin{equation}
\gamma^{0}\gamma^{1}v_{\chi,\sigma}=\chi v_{\chi,\sigma},\ \ v_{\chi,\sigma
}^{\dag}v_{\chi^{\prime},\sigma^{\prime}}=\delta_{\chi,\chi^{\prime}}%
\delta_{\sigma,\sigma^{\prime}}.  \label{de.6}
\end{equation}
The quantum numbers $\chi$ and $\sigma$ describe a spin polarization and
provide parametrization of the solutions. In $d$ dimensions there are exist
only $J_{(d)}=2^{[d/2]-1}$ different spin states. This is a well-known
property related to the specific structure of the projection operator in the
brackets $\left\{ ...\right\} $ in Eq.~(\ref{de.3}) that sets of solutions
of the Dirac equation, which only differ by values of $\chi$ are linearly
dependent. That is why it is sufficient to work only with solutions
corresponding to one of the values of $\chi$; e.g., see Ref. \cite{L-field}
for details.

The scalar functions $\varphi_{n}\left( x\right) $ have to obey the second
order differential equation
\begin{equation}
\left( \hat{p}_{x}^{2}-i\chi\partial_{x}U(x)-\left[ p_{0}-U(x)\right]
^{2}+\pi_{\bot}^{2}\right) \varphi_{n}(x)=0,\ \ \hat{p}_{x}=-i\partial_{x},
\label{de.7}
\end{equation}
where $\pi_{\bot}^{2}=p_{\bot}^{2}+m^{2}$.

%%%%%%%%%%%%%%%%%%%%%%%%%%%%%%%%%%%%%%%%%%

\subsection{Solutions with special left and right asymptotics}

In each interval we introduce new variables\ $\eta_{\mathrm{j}}$, $\mathrm{j}%
=1$ for $x\in\mathrm{I}$, $\mathrm{j}=2$ for $x\in\mathrm{II}$:%
\begin{equation}
\eta_{1}=ih_{1}e^{k_{1}x}\,,\ \ \eta_{2}=ih_{2}e^{-k_{2}x}\,,\ \ h_{\mathrm{j%
}}=\frac{2eE}{k_{\mathrm{j}}^{2}},  \label{de.8}
\end{equation}
and represent the scalar functions $\varphi_{n}\left( x\right) $ as%
\begin{align}
& \varphi_{n}^{\mathrm{j}}(x)=e^{-\eta_{\mathrm{j}}/2}\eta_{\mathrm{j}%
}^{\nu_{\mathrm{j}}}\rho_{\mathrm{j}}(x),  \notag \\
& \nu_{\mathrm{1}}=i\frac{\left\vert p^{\mathrm{L}}\right\vert }{k_{\mathrm{1%
}}},\ \ \left\vert p^{\mathrm{L}}\right\vert =\sqrt{\left[ \pi_{0}\left(
\mathrm{L}\right) \right] ^{2}-\pi_{\bot}^{2}},\ \ \pi _{0}\left( \mathrm{L}%
\right) =p_{0}+\frac{eE}{k_{\mathrm{1}}},  \notag \\
& \nu_{\mathrm{2}}=i\frac{\left\vert p^{\mathrm{R}}\right\vert }{k_{\mathrm{2%
}}},\ \ \left\vert p^{\mathrm{R}}\right\vert =\sqrt{\left[ \pi_{0}\left(
\mathrm{R}\right) \right] ^{2}-\pi_{\bot}^{2}},\ \pi _{0}\left( \mathrm{R}%
\right) =p_{0}-\frac{eE}{k_{\mathrm{2}}}.\ \   \label{de.9}
\end{align}
Here $\pi_{0}\left( \mathrm{L}\right) $ and $\pi_{0}\left( \mathrm{R}\right)
$ are the sum of its asymptotic kinetic and rest energies at $x=-\infty$ and
$x=+\infty$, respectively. We call the quantity $\pi_{\bot}$ the transversal
energy. The functions $\rho_{\mathrm{j}}(x)$ satisfy the confluent
hypergeometric equation \cite{BatE53}
\begin{align}
& \left( \eta_{\mathrm{j}}\frac{d^{2}}{d\eta_{\mathrm{j}}^{2}}+\left[ c_{%
\mathrm{j}}-\eta_{\mathrm{j}}\right] \frac{d}{d\eta_{\mathrm{j}}}-a_{\mathrm{%
j}}\right) \rho_{\mathrm{j}}(x)=0,  \notag \\
& c_{\mathrm{j}}=2\nu_{\mathrm{j}}+1,\ \ a_{1}=\frac{\left( 1-\chi\right) }{2%
}+\nu_{1}-\frac{i\pi_{0}\left( \mathrm{L}\right) }{k_{\mathrm{1}}},  \notag
\\
\ & a_{2}=\frac{\left( 1-\chi\right) }{2}+\nu_{2}+\frac{i\pi_{0}\left(
\mathrm{R}\right) }{k_{\mathrm{2}}}.  \label{de.10}
\end{align}
A fundamental set of solutions for the equation is composed by two linearly
independent confluent hypergeometric functions:%
\begin{equation*}
\Phi\left( a_{\mathrm{j}},c_{\mathrm{j}};\eta_{\mathrm{j}}\right) \ \ \text{%
and}\ \ \eta_{\mathrm{j}}^{1-c_{\mathrm{j}}}e^{\eta_{\mathrm{j}}}\Phi\left(
1-a_{\mathrm{j}},2-c_{\mathrm{j}};-\eta_{\mathrm{j}}\right) ,
\end{equation*}
where
\begin{equation}
\Phi\left( a,c;\eta\right) =1+\frac{a}{c}\frac{\eta}{1!}+\frac{a\left(
a+1\right) }{c\left( c+1\right) }\frac{\eta^{2}}{2!}+\ldots\,.  \label{de.11}
\end{equation}
The general solution of Eq.~(\ref{de.7}) in the intervals $\mathrm{I}$ and $%
\mathrm{II}$ can be expressed as the following linear superposition:
\begin{align}
& \varphi_{n}^{\mathrm{j}}(x)=A_{2}^{\mathrm{j}}y_{1}^{\mathrm{j}}\left(
\eta_{\mathrm{j}}\right) +A_{1}^{\mathrm{j}}y_{2}^{\mathrm{j}}\left( \eta_{%
\mathrm{j}}\right) ,  \notag \\
& y_{1}^{\mathrm{j}}=e^{-\eta_{\mathrm{j}}/2}\eta_{\mathrm{j}}^{\nu _{%
\mathrm{j}}}\Phi\left( a_{\mathrm{j}},c_{\mathrm{j}};\eta_{\mathrm{j}%
}\right) ,  \notag \\
& y_{2}^{\mathrm{j}}=e^{\eta_{_{\mathrm{j}}}/2}\eta_{_{\mathrm{j}}}^{-\nu_{%
\mathrm{j}}}\Phi\left( 1-a_{\mathrm{j}},2-c_{\mathrm{j}};-\eta_{\mathrm{j}%
}\right) =e^{-\eta_{_{\mathrm{j}}}/2}\eta_{_{\mathrm{j}}}^{-\nu_{\mathrm{j}%
}}\Phi\left( a_{\mathrm{j}}-c_{\mathrm{j}}+1,2-c_{\mathrm{j}};\eta_{\mathrm{j%
}}\right) ,  \label{de.12}
\end{align}
with constants $A_{1}^{\mathrm{j}}$ and $A_{2}^{\mathrm{j}}$ being fixed by
boundary conditions.

The complete set of solutions for Klein-Gordon equation can be formally
obtained by setting $\chi\ $equal to zero in all formulas,%
\begin{equation}
\phi_{n}^{\mathrm{j}}(x)=\exp\left[ -ip_{0}t+i\mathbf{p}_{\bot}\mathbf{r}%
_{\bot}\right] \varphi_{n}^{\mathrm{j}}(x).  \label{de.13}
\end{equation}
In this case $n=\mathbf{p}$.

The Wronskian of the $y_{1,2}^{\mathrm{j}}\left( \eta_{\mathrm{j}}\right) $
functions is
\begin{equation}
W=y_{1}^{\mathrm{j}}\frac{d}{d\eta_{\mathrm{j}}}y_{2}^{\mathrm{j}}-y_{2}^{%
\mathrm{j}}\frac{d}{d\eta_{\mathrm{j}}}y_{1}^{\mathrm{j}}=\frac{1-c_{\mathrm{%
j}}}{\eta_{\mathrm{j}}}\,.  \label{de.14}
\end{equation}

In what follows, we use solutions of the Dirac equation denoted as $%
_{\;\zeta }\psi_{n}\left( X\right) $ and $^{\;\zeta}\psi_{n}\left( X\right)
,\ \zeta=\pm\ ,$ with special left and right asymptotics at $x\rightarrow
-\infty$ and $x\rightarrow+\infty$, respectively, where there is no electric
field. Nontrivial solutions $\ ^{\zeta}\psi_{n}\left( X\right) $ exist only
for quantum numbers $n$ that obey the relation%
\begin{equation}
\left[ \pi_{0}\left( \mathrm{R}\right) \right] ^{2}>\pi_{\bot}^{2}%
\Longleftrightarrow\left\{
\begin{array}{l}
\pi_{0}\left( \mathrm{R}\right) >\pi_{\bot} \\
\pi_{0}\left( \mathrm{R}\right) <-\pi_{\bot}%
\end{array}
\right. ,  \label{cond-R}
\end{equation}
whereas nontrivial solutions $_{\zeta}\psi_{n}\left( X\right) $ exist only
for quantum numbers $n$ that obey the relation%
\begin{equation}
\left[ \pi_{0}\left( \mathrm{L}\right) \right] ^{2}>\pi_{\bot}^{2}%
\Longleftrightarrow\left\{
\begin{array}{l}
\pi_{0}\left( \mathrm{L}\right) >\pi_{\bot} \\
\pi_{0}\left( \mathrm{L}\right) <-\pi_{\bot}%
\end{array}
\right. .  \label{cond-L}
\end{equation}
Such solutions have the form (\ref{de.3}) with the functions $\varphi
_{n}\left( x\right) $ denoted as $_{\;\zeta}\varphi_{n}\left( x\right) $ or $%
^{\;\zeta}\varphi_{n}\left( x\right) $, respectively. The latter functions
satisfy Eq.~(\ref{de.7}) and the following asymptotic conditions:%
\begin{equation}
\ _{\zeta}\varphi_{n}(x)=\ _{\zeta}\mathcal{N}e^{i\zeta\left\vert p^{\mathrm{%
L}}\right\vert x}\mathrm{\ \ if\ }\ x\rightarrow-\infty ,\ \
^{\zeta}\varphi_{n}\left( x\right) =\ ^{\zeta}\mathcal{N}e^{i\zeta\left\vert
p^{\mathrm{R}}\right\vert x}\mathrm{\ \ if}\ \ x\rightarrow+\infty\,,\ \
\zeta=\pm.  \label{de.15}
\end{equation}
The solutions $_{\;\zeta}\psi_{n}\left( X\right) $ and $^{\;\zeta}\psi
_{n}\left( X\right) $ asymptotically describe particles with given momenta $%
\zeta\left\vert p^{\mathrm{L}}\right\vert $ and $\zeta\left\vert p^{\mathrm{R%
}}\right\vert $, correspondingly, along the axis $x$.

We consider our theory in a large spacetime box that has a spatial volume $%
V_{\bot}=\prod\limits_{j=2}^{D}K_{j}$ and the time dimension $T$, where all $%
K_{j}$ and $T$ are macroscopically large. The integration over the
transverse coordinates is fulfilled from $-K_{j}/2$ to $+K_{j}/2$, and over
the time $t$ from $-T/2$ to $+T/2$. The limits $K_{j}\rightarrow\infty$ and $%
T\rightarrow \infty$ are assumed in final expressions. In this case the
electric current of the Dirac field through the hypersurface $x=\mathrm{const%
}$,%
\begin{equation}
\left( \psi,\psi^{\prime}\right) _{x}=\int\psi^{\dag}\left( X\right)
\gamma^{0}\gamma^{1}\psi^{\prime}\left( X\right) dtd\mathbf{r}_{\bot}\ ,
\label{c3}
\end{equation}
is $x$-independent. Using Eq.~(\ref{c3}) we subject the solutions $_{\zeta
}\psi_{n}\left( X\right) $ and $^{\zeta}\psi_{n}\left( X\right) $ to the
orthonormality conditions and calculate the normalization constants $_{\zeta
}\mathcal{N}$ and $^{\zeta}\mathcal{N}$\ in (\ref{de.15}) as (see Ref. \cite%
{x-case} for details)%
\begin{align}
_{\zeta}\mathcal{N} & =\ _{\zeta}CY,\mathcal{\ \ }^{\zeta}\mathcal{N}=\
^{\zeta}CY,\mathcal{\ \ }Y=\left( V_{\bot}T\right) ^{-1/2},\ \   \notag \\
_{\zeta}C & =\left[ 2\left\vert p^{\mathrm{L}}\right\vert \left\vert
\pi_{0}\left( \mathrm{L}\right) -\chi\zeta\left\vert p^{\mathrm{L}%
}\right\vert \right\vert \right] ^{-1/2},\ \ ^{\zeta}C=\left[ 2\left\vert p^{%
\mathrm{R}}\right\vert \left\vert \pi_{0}\left( \mathrm{R}\right)
-\chi\zeta\left\vert p^{\mathrm{R}}\right\vert \right\vert \right] ^{-1/2}.\
\label{de.16}
\end{align}

By virtue of these properties, electron (positron) states can be selected as
follows:
\begin{align}
\ _{+}\varphi_{n}\left( x\right) & =\;_{+}\mathcal{N}\exp\left( -i\pi
\nu_{1}/2\right) y_{1}^{\mathrm{1}}\,,\,\ _{-}\varphi_{n}\left( x\right)
=\;_{-}\mathcal{N}\exp\left( i\pi\nu_{1}/2\right) y_{2}^{\mathrm{1}}\,,\ \
x\in\mathrm{I};  \notag \\
\ ^{+}\varphi_{n}\left( x\right) & =\;^{+}\mathcal{N}\exp\left( i\pi
\nu_{2}/2\right) y_{2}^{\mathrm{2}}\,,\,\ ^{-}\varphi_{n}\left( x\right)
=\;^{-}\mathcal{N}\exp\left( -i\pi\nu_{2}/2\right) y_{1}^{\mathrm{2}}\,,\ \
x\in\mathrm{II}.  \label{de.17}
\end{align}
The solutions $^{\zeta}\psi_{n}\left( X\right) $ and $\ _{\zeta}\psi
_{n}\left( X\right) $ are connected by the decomposition%
\begin{align}
\eta_{\mathrm{L}}\ ^{\zeta}\psi_{n}\left( X\right) & =\ _{+}\psi _{n}\left(
X\right) g\left( _{+}\left\vert ^{\zeta}\right. \right) -\
_{-}\psi_{n}\left( X\right) g\left( _{-}\left\vert ^{\zeta}\right. \right) ,
\notag \\
\eta_{\mathrm{R}}\ _{\zeta}\psi_{n}\left( X\right) & =\ ^{+}\psi _{n}\left(
X\right) g\left( ^{+}\left\vert _{\zeta}\right. \right) -\
^{-}\psi_{n}\left( X\right) g\left( ^{-}\left\vert _{\zeta}\right. \right) ,
\label{de.A2}
\end{align}
if the conditions (\ref{cond-R}) and (\ref{cond-L}) are simultaneously
fulfilled, where $\eta_{\mathrm{L/R}}=\mathrm{sgn\ }\pi_{0}\left( \mathrm{L/R%
}\right) $, and the coefficients $g$ are defined by the corresponding inner
product,
\begin{equation}
g\left( \ _{\zeta}\left\vert ^{\zeta^{\prime}}\right. \right) =\left( \
_{\zeta}\psi_{n},\ ^{\zeta^{\prime}}\psi_{n}\right) _{x}\ ,\ \ g\left( \
_{\zeta}\left\vert ^{\zeta^{\prime}}\right. \right) ^{\ast}=g\left(
^{\zeta^{\prime}}\left\vert _{\zeta}\right. \right) .  \label{de.A1a}
\end{equation}
These coefficients satisfy the following unitary relations%
\begin{align}
& \left\vert g\left( _{-}\left\vert ^{+}\right. \right) \right\vert
^{2}=\left\vert g\left( _{+}\left\vert ^{-}\right. \right) \right\vert
^{2},\;\left\vert g\left( _{+}\left\vert ^{+}\right. \right) \right\vert
^{2}=\left\vert g\left( _{-}\left\vert ^{-}\right. \right) \right\vert
^{2},\;\frac{g\left( _{+}\left\vert ^{-}\right. \right) }{g\left(
_{-}\left\vert ^{-}\right. \right) }=\frac{g\left( ^{+}\left\vert
_{-}\right. \right) }{g\left( ^{+}\left\vert _{+}\right. \right) },  \notag
\\
& \left\vert g\left( _{+}\left\vert ^{-}\right. \right) \right\vert
^{2}-\left\vert g\left( _{+}\left\vert ^{+}\right. \right) \right\vert
^{2}=-\eta_{\mathrm{L}}\eta_{\mathrm{R}}.  \label{UR}
\end{align}

Taking into account the complete set of exact solutions (\ref{de.12}) and
mutual decompositions (\ref{de.A2}), for example, one can present the
functions$\ _{-}\varphi_{n}\left( x\right) $ and$\ ^{+}\varphi_{n}\left(
x\right) $ in the form%
\begin{align}
\ \ ^{+}\varphi_{n}\left( x\right) & =\left\{
\begin{array}{l}
\eta_{\mathrm{L}}\left[ \ _{+}\varphi_{n}\left( x\right) g\left(
_{+}|^{+}\right) -\ _{-}\varphi_{n}\left( x\right) g\left( _{-}|^{+}\right) %
\right] \,,\ \ x\in\mathrm{I} \\
\;^{+}\mathcal{N}\exp\left( i\pi\nu_{2}/2\right) y_{2}^{\mathrm{2}}\,,\ \ \
\ \ \ \ \ \ \ \ \ \ \ \ \ \ \ \ \ \ \ \ \ \ \ \,x\in\mathrm{II}%
\end{array}
\right. \,,  \label{de.18} \\
\ \ _{-}\varphi_{n}\left( x\right) & =\left\{
\begin{array}{l}
\;_{-}\mathcal{N}\exp\left( i\pi\nu_{1}/2\right) y_{2}^{\mathrm{1}},\ \ \ \
\ \ \ \ \ \ \ \ \ \ \ \ \ \ \ \ \ \ \ \ \ \ \ \,x\in\mathrm{I} \\
\eta_{\mathrm{R}}\left[ \ ^{+}\varphi_{n}\left( x\right) g\left(
^{+}|_{-}\right) -\ ^{-}\varphi_{n}\left( x\right) g\left( ^{-}|_{-}\right) %
\right] ,\ \ x\in\mathrm{II}%
\end{array}
\right. \,,  \label{de.19}
\end{align}
for the whole axis $x$.

%%%%%%%%%%%%%%%%%%%%%%%%%%%%%%%%%%%%%%%%%%%%%%%%%%%%%%%%%

\subsection{In- and out-sets}

According to the general theory, in the case of $x$-electric potential
steps, the manifold of all the quantum numbers $n$ denoted by $\Omega$ can
be divided into five ranges of quantum numbers $\Omega_{i}$, $i=1,...,5$,
where the corresponding solutions of the Dirac equation have similar forms,
so that $\Omega=\Omega_{1}\cup\cdots\cup\ \Omega_{5}$; see Fig. \ref{fig2}.
Note that the range $\Omega_{3}$ exists if $2\pi_{\bot}<$ $\mathbb{U}$. We
denote\emph{\ }the quantum numbers in corresponding zone $\Omega_{i}$ by $%
n_{i}$. The conditions (\ref{cond-R}) and (\ref{cond-L}) are simultaneously
fulfilled for $\Omega_{i}$, $i=1,3,5$, as follows%
\begin{align}
\pi_{0}\left( \mathrm{L}\right) & >\pi_{0}\left( \mathrm{R}\right)
>\pi_{\bot}\;\mathrm{if}\;n\in\Omega_{1},\;\;\pi_{0}\left( \mathrm{R}\right)
<\pi_{0}\left( \mathrm{L}\right) <-\pi_{\bot}\;\mathrm{if}\;n\in\Omega _{5},
\notag \\
\pi_{0}\left( \mathrm{L}\right) & >\pi_{\bot},\;\pi_{0}\left( \mathrm{R}%
\right) <-\pi_{\bot}\;\mathrm{if}\;n\in\Omega_{3}.  \label{Omega}
\end{align}
For the detailed description of the ranges $\Omega_{i}$ and their properties
see Ref. \cite{x-case}.

The exact expressions for $g$'s can be obtained from Eqs. (\ref{de.18}) and (%
\ref{de.19}) as follows. The functions$\ _{-}\varphi_{n}\left( x\right) $
and $\ ^{+}\varphi_{n}\left( x\right) $ given by Eqs.~(\ref{de.18}) and (\ref%
{de.19}) and their derivatives satisfy the following gluing conditions:
\begin{equation}
\left. \ _{-}^{+}\varphi_{n}(x)\right\vert _{x=-0}=\left. \
_{-}^{+}\varphi_{n}(x)\right\vert _{x=+0}\,,\ \ \left. \partial_{x}\
_{-}^{+}\varphi_{n}(x)\right\vert _{x=-0}=\left. \partial_{x}\
_{-}^{+}\varphi _{n}(x)\right\vert _{x=+0}\,.  \label{de.20}
\end{equation}
Using Eq.~(\ref{de.20}) and the Wronskian (\ref{de.14}), one can find each
coefficient $g\left( _{\zeta}|^{\zeta^{\prime}}\right) $ and $g\left(
^{\zeta}|_{\zeta^{\prime}}\right) $ in Eqs.~(\ref{de.18}) and (\ref{de.19}).
For example, applying these conditions to the set (\ref{de.18}), one can
find the coefficient $g\left( _{-}|^{+}\right) $:%
\begin{align}
& \ g\left( _{-}|^{+}\right) =C\Delta,\ \ C=\frac{\eta_{\mathrm{L}}}{2}\sqrt{%
\frac{\left\vert \pi_{0}\left( \mathrm{L}\right) +\chi\left\vert p^{\mathrm{L%
}}\right\vert \right\vert }{\left\vert p^{\mathrm{L}}\right\vert \left\vert
p^{\mathrm{R}}\right\vert \left\vert \pi_{0}\left( \mathrm{R}\right)
-\chi\left\vert p^{\mathrm{R}}\right\vert \right\vert }}\exp\left[ \frac{i\pi%
}{2}\left( \nu_{\mathrm{2}}-\nu_{\mathrm{1}}\right) \right] ,  \notag \\
& \ \Delta=\left. \left[ k_{\mathrm{1}}h_{\mathrm{1}}y_{2}^{\mathrm{2}}\frac{%
d}{d\eta_{\mathrm{1}}}y_{1}^{\mathrm{1}}+k_{\mathrm{2}}h_{\mathrm{2}}y_{1}^{%
\mathrm{1}}\frac{d}{d\eta_{\mathrm{2}}}y_{2}^{\mathrm{2}}\right] \right\vert
_{x=0}.  \label{de.21}
\end{align}
The same can be done to Eq.~(\ref{de.19}) to obtain%
\begin{align}
& \ g\left( ^{+}|_{-}\right) =C^{\prime}\Delta^{\prime},\ \ C^{\prime }=%
\frac{-\eta_{\mathrm{R}}}{2}\sqrt{\frac{\left\vert \pi_{0}\left( \mathrm{R}%
\right) -\chi\left\vert p^{\mathrm{R}}\right\vert \right\vert }{\left\vert
p^{\mathrm{R}}\right\vert \left\vert p^{\mathrm{L}}\right\vert \left\vert
\pi_{0}\left( \mathrm{L}\right) +\chi\left\vert p^{\mathrm{L}}\right\vert
\right\vert }}\exp\left[ \frac{i\pi}{2}\left( \nu_{\mathrm{1}}-\nu_{\mathrm{2%
}}\right) \right] ,  \notag \\
& \ \Delta^{\prime}=\left. \left[ k_{\mathrm{1}}h_{\mathrm{1}}y_{1}^{\mathrm{%
2}}\frac{d}{d\eta_{\mathrm{1}}}y_{2}^{\mathrm{1}}+k_{\mathrm{2}}h_{\mathrm{2}%
}y_{2}^{\mathrm{1}}\frac{d}{d\eta_{\mathrm{2}}}y_{1}^{\mathrm{2}}\right]
\right\vert _{x=0}.  \label{de.22}
\end{align}
One can easily verify that the symmetry under a simultaneous change $k_{%
\mathrm{1}}\leftrightarrows k_{\mathrm{2}}$ and $\pi_{0}\left( \mathrm{L}%
\right) \leftrightarrows-\pi_{0}\left( \mathrm{R}\right) $ holds,%
\begin{equation}
g\left( ^{+}|_{-}\right) \leftrightarrows-\eta_{\mathrm{L}}\eta_{\mathrm{R}%
}g\left( _{-}|^{+}\right) \,.  \label{de.23}
\end{equation}

A formal transition to the Klein-Gordon case can be done by setting$\ \chi
=0 $ and $\eta _{\mathrm{L}}=\eta _{\mathrm{R}}=1$ in Eqs.~(\ref{de.21}) and
(\ref{de.22}). In this case, normalization factors $\ _{\zeta }C$ and $%
^{\zeta }C$ are%
\begin{equation}
_{\zeta }C=\left\vert 2p^{\mathrm{L}}\right\vert ^{-1/2},\ \ ^{\zeta
}C=\left\vert 2p^{\mathrm{R}}\right\vert ^{-1/2}.  \label{de.24}
\end{equation}%
The coefficient $g\left( _{-}|^{+}\right) $ for scalar particles is
\begin{equation}
\ g\left( _{-}|^{+}\right) =C_{\mathrm{sc}}\left. \Delta \right\vert _{\chi
=0},\ \ C_{\mathrm{sc}}=\sqrt{\frac{1}{4\left\vert p^{\mathrm{L}}\right\vert
\left\vert p^{\mathrm{R}}\right\vert }}\exp \left[ \frac{i\pi }{2}\left( \nu
_{\mathrm{2}}-\nu _{\mathrm{1}}\right) \right]  \label{de.25}
\end{equation}%
with $\Delta $ given by Eq.~(\ref{de.21}) . The symmetry under the
simultaneous change $k_{\mathrm{1}}\leftrightarrows k_{\mathrm{2}}$ and $\pi
_{0}\left( \mathrm{L}\right) \leftrightarrows -\pi _{0}\left( \mathrm{R}%
\right) $ holds as\
\begin{equation}
g\left( ^{+}|_{-}\right) \leftrightarrows g\left( _{-}|^{+}\right) \,.
\label{de.26}
\end{equation}

As follows from Eqs.~(\ref{de.21}), (\ref{de.22}), and (\ref{de.25}), if
either $\left\vert p^{\mathrm{R}}\right\vert $ or $\left\vert p^{\mathrm{L}%
}\right\vert $ tends to zero, one of the following limits holds true:%
\begin{equation}
\left\vert g\left( _{-}|^{+}\right) \right\vert ^{-2}\sim\left\vert p^{%
\mathrm{R}}\right\vert \rightarrow0,\ \ \left\vert g\left( ^{+}|_{-}\right)
\right\vert ^{-2}\sim\left\vert p^{\mathrm{L}}\right\vert \rightarrow0,\ \
\forall\lambda\neq0.  \label{de.29}
\end{equation}
These properties are essential for the justification of \textrm{in}- and
\textrm{out}-particle interpretation in the general construction \cite%
{x-case}.

However, it should be noted that quantum field theory deals with physical
quantities that are presented by volume integrals on $t$-constant
hyperplane. The time-independent inner product for any pair of solutions of
the Dirac equation, $\psi _{n}\left( X\right) $ and $\psi _{n^{\prime
}}^{\prime }\left( X\right) $, is defined on the $t=$\textrm{const}
hyperplane as follows:
\begin{equation}
\left( \psi _{n},\psi _{n^{\prime }}^{\prime }\right) =\int_{V_{\bot }}d%
\mathbf{r}_{\bot }\int\limits_{-K^{\left( \mathrm{L}\right) }}^{K^{\left(
\mathrm{R}\right) }}\psi _{n}^{\dag }\left( X\right) \psi _{n^{\prime
}}^{\prime }\left( X\right) dx,\ \   \label{t4}
\end{equation}%
where the improper integral over $x$ in the right-hand side of Eq. (\ref{t4}%
) is reduced to its special principal value to provide a certain additional
property important for us and the limits $K^{\left( \mathrm{L}/\mathrm{R}%
\right) }\rightarrow \infty $ are assumed in final expressions. As a result,
we can see that all the wave functions having different quantum numbers $n$
are orthogonal with respect of the inner product (\ref{t4}). We can find the
linear independent pairs of $\psi _{n}\left( X\right) $ and $\psi
_{n^{\prime }}^{\prime }\left( X\right) $ for each $n$ and identify initial
and final states on the $t=$\textrm{const} hyperplane as follows (see Ref.
\cite{x-case} for details):%
\begin{align}
& \mathrm{in-solutions:\ }_{+}\psi _{n_{1}},\ ^{-}\psi _{n_{1}};\mathrm{\ }%
_{-}\psi _{n_{5}},\ ^{+}\psi _{n_{5}};\ \ _{-}\psi _{n_{3}},\ ^{-}\psi
_{n_{3}}\ ,  \notag \\
& \mathrm{out-solutions:\ }_{-}\psi _{n_{1}},\ ^{+}\psi _{n_{1}};\ _{+}\psi
_{n_{5}},\ ^{-}\psi _{n_{5}};\ \ _{+}\psi _{n_{3}},\ ^{+}\psi _{n_{3}}.
\label{in-out}
\end{align}%
In the ranges $\Omega _{2}$ ($-\pi _{\bot }<\pi _{0}\left( \mathrm{R}\right)
<\pi _{\bot }$ and $\pi _{0}\left( \mathrm{L}\right) >\pi _{\bot }$) any
solution has zero right asymptotic, which means that we deal with electron
standing waves $\psi _{n_{2}}\left( X\right) $. It means that we deal with a
total reflection.\textbf{\ }Similarly, we can treat positron standing waves $%
\psi _{n_{4}}\left( X\right) $ in the range $\Omega _{4}$ ($-\pi _{\bot
}<\pi _{0}\left( \mathrm{L}\right) <\pi _{\bot }$ and $\pi _{0}\left(
\mathrm{R}\right) <-\pi _{\bot }$) and see a total reflection of positrons.
It has to be noted that the complete set of in- and out-solution must
include solution $\psi _{n_{2}}\left( X\right) $ and $\psi _{n_{4}}\left(
X\right) $.

Using the identification (\ref{in-out}) we decompose the Heisenberg field
operator $\hat{\Psi}\left( X\right) $ in two sets of solutions of the Dirac
equation (\ref{de.1}) complete on the $t=\mathrm{const}$ hyperplane.
Operator-valued coefficients in such decompositions are creation and
annihilation operators of electrons and positrons which do not depend on
coordinates. Following this way we complete initial and final sets of
creation and annihilation operators as

\begin{align}
& \mathrm{in-set:\ }_{+}a_{n_{1}}(\mathrm{in}),\ \ ^{-}a_{n_{1}}(\mathrm{in}%
);\mathrm{\ }\ _{-}b_{n_{5}}(\mathrm{in}),\ \ ^{+}b_{n_{5}}(\mathrm{in});\ \
\ _{-}b_{n_{3}}(\mathrm{in}),\ \ ^{-}a_{n_{3}}(\mathrm{in})\ ,  \notag \\
& \mathrm{out-set:\ }\ _{-}a_{n_{1}}(\mathrm{out}),\ \ ^{+}a_{n_{1}}(\mathrm{%
out});\ _{+}b_{n_{5}}(\mathrm{out}),\ \ ^{-}b_{n_{5}}(\mathrm{out});\
_{+}b_{n_{3}}(\mathrm{out}),\ \ ^{+}a_{n_{3}}(\mathrm{out}).  \label{sets}
\end{align}
We interpret all $a$ and $b\ $as annihilation and all $a^{\dag}$ and $%
b^{\dag }$ as creation operators. All $a$ and $a^{\dag}$ are interpreted\ as
describing electrons and all $b$ and $b^{\dag}$ as describing positrons. All
the operators labeled by the argument \textrm{in} are interpreted\ as
\textrm{in}-operators, whereas all the operators labeled by the argument
\textrm{out} as \textrm{out}-operators. This identification is confirmed by
a detailed mathematical and physical analysis of solutions of the Dirac
equation with subsequent QFT analysis of correctness of such an
identification in Ref. \cite{x-case}. We define two vacuum vectors $%
\left\vert 0,\mathrm{in}\right\rangle $ and $\left\vert 0,\mathrm{out}%
\right\rangle $, one of which is the\ zero-vector for all \textrm{in}%
-annihilation operators and the other is zero-vector for all $\mathrm{out}$%
-annihilation operators. Besides, both vacua are zero-vectors for the
annihilation operators $a_{n_{2}}$ and $b_{n_{4}}$. We know that in the
ranges $\Omega_{i}$, $i=1,2,4,5$ the partial vacua, $\left\vert 0,\mathrm{in}%
\right\rangle ^{\left( i\right) }$ and $\left\vert 0,\mathrm{out}%
\right\rangle ^{\left( i\right) }$,$\ $ are stable. The vacuum-to-vacuum
transition amplitude $c_{v}$ coincides with the vacuum-to-vacuum transition
amplitude $c_{v}^{\left( 3\right) }$ in the Klein zone $\Omega_{3}$,%
\begin{equation}
c_{v}=\langle0,\mathrm{out}|0,\mathrm{in}\rangle=c_{v}^{\left( 3\right) }=\
^{\left( 3\right) }\langle0,\mathrm{out}|0,\mathrm{in}\rangle^{(3)}\ .
\label{de.A10}
\end{equation}

%%%%%%%%%%%%%%%%%%%%%%%%%%%%%%%%%%%%%%%%%%

\section{Scattering and reflection of particles outside of the Klein zone
\label{scattering}}

To extract results of the one-particle scattering theory, all the
constituent quantities, such as reflection and transmission coefficients
etc., have to be represented with the help of the mutual decomposition
coefficients $g$.

As an example, in the range $\Omega_{1}$, one can calculate absolute $\tilde{%
R}$ and relative $R$ amplitudes of an electron reflection, and absolute $%
\tilde{T}$ and relative $T$ amplitudes of an electron transmission, which
can be presented as the following matrix elements%
\begin{align}
& \ R_{+,n}=\tilde{R}_{+,n}c_{v}^{-1}\ \ \ \tilde{R}_{+,n}=\langle 0,\mathrm{%
out}|\ _{-}a_{n}\left( \mathrm{out}\right) \ _{+}a_{n}^{\dag }\left( \mathrm{%
in}\right) |0,\mathrm{in}\rangle,  \notag \\
& \ T_{+,n}=\tilde{T}_{+,n}c_{v}^{-1}\ \ \ \tilde{T}_{+,n}=\langle 0,\mathrm{%
out}|\ ^{+}a_{n}\left( \mathrm{out}\right) \ _{+}a_{n}^{\dag }\left( \mathrm{%
in}\right) |0,\mathrm{in}\rangle,  \notag \\
& \ R_{-,n}=\tilde{R}_{-,n}c_{v}^{-1}\ \ \ \tilde{R}_{-,n}=\langle 0,\mathrm{%
out}|\ ^{+}a_{n}\left( \mathrm{out}\right) \ ^{-}a_{n}^{\dag }\left( \mathrm{%
in}\right) |0,\mathrm{in}\rangle,  \notag \\
& \ T_{-,n}=\tilde{T}_{-,n}c_{v}^{-1}\ \ \ \tilde{T}_{-,n}=\langle 0,\mathrm{%
out}|\ _{-}a_{n}\left( \mathrm{out}\right) \ ^{-}a_{n}^{\dag }\left( \mathrm{%
in}\right) |0,\mathrm{in}\rangle,  \label{sc.1}
\end{align}
It follows from the Eq. (\ref{sc.1}) that the relative reflection $%
\left\vert R_{\zeta,n}\right\vert ^{2}$ and transition $\left\vert
T_{\zeta,n}\right\vert ^{2}$ probabilities are
\begin{equation}
\left\vert T_{\zeta,n}\right\vert ^{2}=1-\left\vert R_{\zeta,n}\right\vert
^{2},\ \ \left\vert R_{\zeta,n}\right\vert ^{2}=\left[ 1+\left\vert g\left(
_{-}|^{+}\right) \right\vert ^{-2}\right] ^{-1},\ \ \zeta=\pm.  \label{sc.2}
\end{equation}

Similar expressions can be derived for positron amplitudes in the range $%
\Omega_{5}$. In particular, relation (\ref{sc.2}) holds true literally for
the positrons in the range $\Omega_{5}$. It is clear that $\left\vert
R_{\zeta,n}\right\vert ^{2}\leq1$. This result may be interpreted as QFT
justification of the rules of time-independent potential scattering theory
in the ranges $\Omega_{1}$ and $\Omega_{5}$.

Amplitudes of Klein-Gordon particle reflection and transmission in the
ranges $\Omega_{i}$, $i=1,2,4,5$ have the same form as in the Dirac particle
case with coefficients $g$ given by the corresponding inner product.
Substituting the corresponding coefficients $g$ into relations (\ref{sc.2}),
one can find explicitly reflection and transmission probabilities in the
field under consideration.

It is clear that $\left\vert g\left( _{-}|^{+}\right) \right\vert ^{-2}$ and
then $\left\vert R_{\zeta,n}\right\vert ^{2}$ and $\left\vert T_{\zeta
,n}\right\vert ^{2}$ are functions of modulus squared of transversal
momentum $\mathbf{p}_{\perp}^{2}$. It follows from Eq.~(\ref{de.23}) and
Eq.~(\ref{de.26}), respectively, that $\left\vert R_{\zeta,n}\right\vert
^{2} $ and $\left\vert T_{\zeta,n}\right\vert ^{2}$ are invariant under the
simultaneous change $k_{\mathrm{1}}\leftrightarrows k_{\mathrm{2}}$ and $%
\pi_{0}\left( \mathrm{L}\right) \leftrightarrows-\pi_{0}\left( \mathrm{R}%
\right) $ for both fermions and bosons. Then if $k_{\mathrm{1}}=k_{\mathrm{2}%
}$, $\left\vert R_{\zeta,n}\right\vert ^{2}$ and $\left\vert
T_{\zeta,n}\right\vert ^{2}$ appear to be an even function of $p_{0}$. The
limits (\ref{de.29}) imply that

(i) $\left\vert g\left( _{-}|^{+}\right) \right\vert ^{-2}\rightarrow0$ in
the range $\Omega_{1}$ if $n$ tends to the boundary with the range $\Omega
_{2}$ $\left( \left\vert p^{\mathrm{R}}\right\vert \rightarrow0\right) $;

(ii)$\ \left\vert g\left( _{-}|^{+}\right) \right\vert ^{-2}\rightarrow0$ in
the range $\Omega_{5}$ if $n$ tends to the boundary with the range $\Omega
_{4}$ $\left( \left\vert p^{\mathrm{L}}\right\vert \rightarrow0\right) $.

Thus, in these two cases the relative probabilities of reflection $%
\left\vert R_{\zeta,n}\right\vert ^{2}$ tend to unity; i.e. they are
continuous functions of the quantum numbers $n$ on the boundaries. It can be
also seen that $\left\vert R_{\zeta,n}\right\vert ^{2}\rightarrow0$ as $%
p_{0}\rightarrow \pm\infty$.

%%%%%%%%%%%%%%%%%%%%%%%%%%%%%%%%%%%%%%%%%%

\section{Processes in the Klein zone\label{Klein}}

\subsection{General}

Here we consider possible processes in the Klein zone, $\Omega_{3}$,
following the general consideration \cite{x-case}. It is of special interest
due to the vacuum instability. Due to specific choice of quantum numbers,
processes for different modes $n$ are independent. One sees that physical
quantities are factorized with respect to quantum modes $n$ and calculations
in each mode can be performed separately. In particular, one can represent
the introduced vacua, $\left\vert 0,\mathrm{in}\right\rangle $ and $%
\left\vert 0,\mathrm{out}\right\rangle $, as tensor products of all the
corresponding partial vacua in each mode $n$, respectively, and see that the
probability for a vacuum to remain a vacuum can be expressed as product of
the probabilities $p_{v}^{n}$ for a partial vacuum to remain a vacuum in
each mode $n$,
\begin{equation}
P_{v}=|c_{v}|^{2}=|c_{v}^{\left( 3\right) }|^{2}=\prod\limits_{n\in
\Omega_{3}}p_{v}^{n},  \label{Pv}
\end{equation}
where it is taken into account that in the ranges $\Omega_{i}$, $i=1,2,4,5$
the partial vacua are stable.

The differential mean numbers of electrons and positrons from
electron-positron pairs created are equal:%
\begin{align}
& N_{n}^{a}\left( \mathrm{out}\right) =\left\langle 0,\mathrm{in}\left\vert
\ ^{+}a_{n}^{\dagger}(\mathrm{out})\ ^{+}a_{n}(\mathrm{out})\right\vert 0,%
\mathrm{in}\right\rangle =\left\vert g\left( _{-}\left\vert ^{+}\right.
\right) \right\vert ^{-2},  \notag \\
& N_{n}^{b}\left( \mathrm{out}\right) =\left\langle 0,\mathrm{in}\left\vert
\ _{+}b_{n}^{\dagger}(\mathrm{out})\ _{+}b_{n}(\mathrm{out})\right\vert 0,%
\mathrm{in}\right\rangle =\left\vert g\left( _{+}\left\vert ^{-}\right.
\right) \right\vert ^{-2},  \notag \\
& N_{n}^{\mathrm{cr}}=N_{n}^{b}\left( \mathrm{out}\right) =N_{n}^{a}\left(
\mathrm{out}\right) ,\ \ n\in\Omega_{3},  \label{7.5}
\end{align}
and they present the number of pairs created, $N_{n}^{\mathrm{cr}}$. It
follows from the Eqs.~(\ref{de.21}) and (\ref{de.25}) that%
\begin{align}
N_{n}^{\mathrm{cr}} & =\left\vert C\Delta\right\vert ^{-2}\ \ \ \ \ \ \ \ \
\ \ \text{\textrm{for fermions,}}  \notag \\
N_{n}^{\mathrm{cr}} & =\left\vert C_{\mathrm{sc}}\left. \Delta\right\vert
_{\chi=0}\right\vert ^{-2}\,\text{\ \textrm{for bosons.}}  \label{de.A12}
\end{align}
It is clear that $N_{n}^{\mathrm{cr}}$ is a function of modulus squared of
transversal momentum $\mathbf{p}_{\perp}^{2}$. It follows from Eq.~(\ref%
{de.23}) and Eq.~(\ref{de.26}), respectively, that $N_{n}^{\mathrm{cr}}$ is
invariant under the simultaneous change $k_{\mathrm{1}}\leftrightarrows k_{%
\mathrm{2}}$ and $\pi_{0}\left( \mathrm{L}\right)
\leftrightarrows-\pi_{0}\left( \mathrm{R}\right) $ for both fermions and
bosons. Then if $k_{\mathrm{1}}=k_{\mathrm{2}}$, $N_{n}^{\mathrm{cr}}$
appears to be an even function of $p_{0}$.

From properties (\ref{de.29}), one finds that $N_{n}^{\mathrm{cr}%
}\rightarrow0$ if $n$ tends to the boundary with either the range $%
\Omega_{2} $ ($\left\vert p^{\mathrm{R}}\right\vert \rightarrow0$) or the
range $\Omega _{4}$ ($\left\vert p^{\mathrm{L}}\right\vert \rightarrow0$),

\begin{equation}
N_{n}^{\mathrm{cr}}\sim\left\vert p^{\mathrm{R}}\right\vert \rightarrow 0,\
\ N_{n}^{\mathrm{cr}}\sim\left\vert p^{\mathrm{L}}\right\vert \rightarrow0,\
\ \forall\lambda\neq0;  \label{Nb}
\end{equation}
in the latter ranges, the vacuum is stable.

Absolute values of the asymptotic momenta $\left\vert p^{\mathrm{L}%
}\right\vert $ and $\left\vert p^{\mathrm{R}}\right\vert \ $are determined
by the quantum numbers $p_{0}$ and $\mathbf{p}_{\bot}$, see Eq. (\ref{de.9}%
). This fact imposes certain relation between both quantities. In
particular, one can see that $d\left\vert p^{\mathrm{L}}\right\vert
/d\left\vert p^{\mathrm{R}}\right\vert <0,$ and at any given $\mathbf{p}%
_{\bot}$ these quantities are restricted inside the range $\Omega_{3}$,%
\begin{equation}
0\leq\left\vert p^{\mathrm{R/L}}\right\vert \leq p^{\mathrm{\max}},\;\;p^{%
\mathrm{\max}}=\sqrt{\mathbb{U}\left( \mathbb{U}-2\pi_{\bot}\right) }.
\label{d8}
\end{equation}
It implies that
\begin{equation}
0\leq\left\vert \left\vert p^{\mathrm{L}}\right\vert -\left\vert p^{\mathrm{R%
}}\right\vert \right\vert \leq p^{\mathrm{\max}}.  \label{g8}
\end{equation}
We have $\left\vert p^{\mathrm{L}}\right\vert =k_{\mathrm{1}}\left\vert \nu_{%
\mathrm{1}}\right\vert $, $\left\vert p^{\mathrm{R}}\right\vert =k_{\mathrm{2%
}}\left\vert \nu_{\mathrm{2}}\right\vert $, and $\mathbb{U}=eE\left( k_{%
\mathrm{1}}^{-1}+k_{\mathrm{2}}^{-1}\right) $ for the case under
consideration. Then for any $p_{0}$ and $\mathbf{p}_{\bot}$ the numbers $%
N_{n}^{\mathrm{cr}}$ are negligible if the Klein zone is tiny,%
\begin{equation}
N_{n}^{\mathrm{cr}}\sim\left\vert p^{\mathrm{R}}p^{\mathrm{L}}\right\vert
\rightarrow0\;\;\mathrm{if}\;\;p^{\mathrm{\max}}\rightarrow0.  \label{tiny}
\end{equation}

The total number of pairs $N^{\mathrm{cr}}$ created by the field under
consideration can be calculated by summation over all possible quantum
numbers in the Klein zone. Calculating this number in the fermionic case,
one has to sum the corresponding differential mean numbers $N_{n}^{\mathrm{cr%
}}$ over the spin projections and over the transversal momenta $\mathbf{p}%
_{\bot}$ and energy $p_{0}$. Since the $N_{n}^{\mathrm{cr}}$ do not depend
on the spin polarization parameters $\sigma$, the sum over the spin
projections produces only the factor $J_{(d)}=2^{\left[ d/2\right] -1}$. The
sum over the momenta and the energy transforms into an integral in the
following way:
\begin{equation}
N^{\mathrm{cr}}=\sum_{n\in\Omega_{3}}N_{n}^{\mathrm{cr}}=\sum_{\mathbf{p}%
_{\bot},\ p_{0}\in\Omega_{3}}\sum_{\sigma}N_{n}^{\mathrm{cr}}\rightarrow
\frac{V_{\bot}\ TJ_{(d)}}{\left( 2\pi\right) ^{d-1}}\int_{\Omega_{3}}dp_{0}d%
\mathbf{p}_{\bot}\ N_{n}^{\mathrm{cr}},  \label{dtq.3}
\end{equation}
where $V_{\bot}$ is the spatial volume of the $(d-1)$ dimensional
hypersurface orthogonal to the electric field direction, $x$, and $T$ is the
time duration of the electric field. The total number of bosonic pairs
created in all possible states follows from Eq.~(\ref{dtq.3}) at $J_{(d)}=1$.

Both for fermions and bosons,{\large \ }the relative probabilities of an
electron reflection, a pair creation, and the probability for a partial
vacuum to remain a vacuum in a mode $n$ can be expressed via differential
mean numbers of created pairs $N_{n}^{\mathrm{cr}}$,
\begin{align}
& p_{n}(+|+)=|\langle0,\mathrm{out}|\ ^{+}a_{n}(\mathrm{out})\
^{-}a_{n}^{\dagger}(\mathrm{in})|0,\mathrm{in}\rangle|^{2}P_{v}^{-1}=\left(
1-\kappa N_{n}^{\mathrm{cr}}\right) ^{-1},  \notag \\
& p_{n}(+-|0)=|\langle0,\mathrm{out}|\ ^{+}a_{n}(\mathrm{out})\ _{+}b_{n}(%
\mathrm{out})|0,\mathrm{in}\rangle|^{2}P_{v}^{-1}=N_{n}^{\mathrm{cr}}\left(
1-\kappa N_{n}^{\mathrm{cr}}\right) ^{-1}\;,  \notag \\
& p_{v}^{n}=\left( 1-\kappa N_{n}^{\mathrm{cr}}\right)
^{\kappa},\;\kappa=\left\{
\begin{array}{c}
+1\ \mathrm{for\ fermions} \\
-1\ \mathrm{for\ bosons}%
\end{array}
\right. ,  \label{p}
\end{align}
where $P_{v}$ is defined by Eq.~(\ref{Pv}). The partial absolute
probabilities of an electron reflection and a pair creation in a mode $n$ are%
\begin{equation}
P_{n}(+|+)=p_{n}(+|+)p_{v}^{n},\ \ P_{n}(+-|0)=p_{n}(+-|0)p_{v}^{n},
\label{pabs}
\end{equation}
respectively. The relative probabilities for a positron reflection $%
p_{n}(-|-)$ and a pair annihilation $p_{n}\left( 0|-+\right) $ coincide with
the probabilities $p_{n}(+|+)$ and $p_{n}(+-|0)$, respectively.

We recall, as it follows from the general consideration \cite{x-case}, that
if there exists an \textrm{in}-particle in the Klein zone, it will be
subjected to the total\emph{\ }reflection. For example, it can be
illustrated by a result following from Eqs.~(\ref{p}) and (\ref{pabs}), the
probability of reflection of a Dirac particle with given quantum numbers $n$%
, under the condition that all other partial vacua remain vacua, is $%
P_{n}(+|+)=1$. In the Dirac case, the presence of an \textrm{in}-particle
with a given $n\in \Omega_{3}$ disallows the pair creation from the vacuum
in this state due to the Pauli principle. By the same reason, if an initial
state is vacuum, there are only two possibilities in a cell of the space
with given quantum number $n$\emph{,} namely, this partial vacuum remains a
vacuum, or with the probability $P_{n}(+-|0)$ a pair with the quantum number
$n$ will be created. It is in agreement with the probability conservation
law $p_{v}^{n}+P_{n}(+-|0)=1$ that follows from Eqs.~(\ref{p}) and (\ref%
{pabs}).

Of course, pairs of bosons can be created from the vacuum in any
already-occupied states. For example, the conditional probability of a pair
creation with a given quantum numbers $n$, under the condition that all
other partial vacua with the quantum numbers\emph{\ }$m\neq n$ remain the
vacua is the sum of probabilities of creation from vacuum for any number $l$
of pairs%
\begin{equation}
P_{n}(\mathrm{pairs}|0)=p_{v}^{n}\sum_{l=1}^{\infty}p_{n}(+-|0)^{l}.
\label{r13}
\end{equation}
In this case the probability conservation law has the form of a sum of
probabilities of all possible events in a cell of the space of quantum
numbers $n$:%
\begin{equation}
P(\mathrm{pairs}|0)_{n}+p_{v}^{n}=1.  \label{r14}
\end{equation}

%%%%%%%%%%%%%%%%%%%%%%%%%%%%%%%%%%%%%%%%%%%%

\subsection{Small-gradient field\label{slow_field}}

The inverse parameters $k_{\mathrm{1}}^{-1}$, $k_{\mathrm{2}}^{-1}$
represent scales of growth and decay of the electric field in the intervals $%
\mathrm{I}$ and $\mathrm{II}$, respectively. In particular, we have a
small-gradient field at small values of both $k_{\mathrm{1}},k_{\mathrm{2}%
}\rightarrow0$, obeying the conditions%
\begin{equation}
\min\left( h_{\mathrm{1}},h_{\mathrm{2}}\right) \gg\max\left(
1,m^{2}/eE\right) \,.  \label{svf.1}
\end{equation}
This case can be considered as a two-parameter regularization for an uniform
electric field. That is a reason the Klein zone, $\Omega_{3}$, is of
interest under the condition (\ref{svf.1}).

Let us analyze how the numbers $N_{n}^{\mathrm{cr}}$ depend on the
parameters $p_{0}$ and $\pi_{\perp}$. By virtue of the symmetry properties
of $N_{n}^{\mathrm{cr}}$ discussed above, one can only consider $p_{0}$
either{\Huge \ }positive or negative.

Let us, for example, consider the interval of negative energies $p_{0}\leq0$%
. \ In this case, taking into account that both $\pi_{0}\left( \mathrm{L}%
\right) $ and $\pi_{0}\left( \mathrm{R}\right) $ satisfy the inequalities
given by Eq.~(\ref{Omega}) in the range $\Omega_{3}$, we see that $\pi
_{0}\left( \mathrm{L}\right) $ varies greatly while $\pi_{0}\left( \mathrm{R}%
\right) $ is negative and very large,
\begin{equation}
\pi_{\perp}\leq\pi_{0}\left( \mathrm{L}\right) \leq\frac{eE}{k_{\mathrm{1}}}%
,\;\frac{eE}{k_{\mathrm{2}}}\leq-\pi_{0}\left( \mathrm{R}\right) \leq \frac{%
eE}{k_{\mathrm{2}}}+\frac{eE}{k_{\mathrm{1}}}-\pi_{\perp}.  \label{svf.5}
\end{equation}
It can be seen from the asymptotic behavior of a confluent hypergeometric
function that $N_{n}^{\mathrm{cr}}$ is exponentially small, $N_{n}^{\mathrm{%
cr}}\lesssim\exp\left[ -2\sqrt{h_{2}\left\vert \pi_{0}\left( \mathrm{R}%
\right) \right\vert /k_{2}}\right] $, if $\left\vert p^{\mathrm{R}%
}\right\vert \ll\left\vert \pi_{0}\left( \mathrm{R}\right) \right\vert $ for
large $\left\vert \pi_{0}\left( \mathrm{R}\right) \right\vert $. In this
case, $\pi_{\perp}\sim eEk_{\mathrm{2}}^{-1}$. Then the range of fixed $%
\pi_{\perp}$ is of interest, and in the following we assume that condition%
\begin{equation}
\sqrt{\lambda}<K_{\bot},\ \lambda=\frac{\pi_{\bot}^{2}}{eE}  \label{svf.3}
\end{equation}
holds true, where any given number $K_{\bot}$ satisfies the inequality%
\begin{equation}
\min\left( h_{\mathrm{1}},h_{\mathrm{2}}\right) \gg K_{\bot}^{2}\gg
\max\left( 1,m^{2}/eE\right) \,.  \label{svf.2}
\end{equation}

Using the asymptotic expressions of the confluent hypergeometric functions
we find that the differential mean numbers of created pairs $N_{n}^{\mathrm{%
cr}}$, given by Eq.~(\ref{de.A12}), can be approximated by the forms (\ref%
{svf.11}); see details in Appendix \ref{App.1}. These forms are
exponentially small if $\pi _{0}\left( \mathrm{L}\right) \sim \pi _{\perp }$%
. Then substantial value of $N_{n}^{\mathrm{cr}}$ are formed in the range%
\begin{equation}
h_{\mathrm{1}}\geq 2\pi _{0}\left( \mathrm{L}\right) /k_{1}>K\gg K_{\bot },
\label{svf.12b}
\end{equation}%
where \ $K$ is any given number and $K_{\bot }$ satisfies the inequalities (%
\ref{svf.3}) and (\ref{svf.2}). In this range we approximate the
distributions (\ref{svf.11}) by the formula%
\begin{equation}
N_{n}^{\mathrm{cr}}\approx \exp \left\{ -\frac{2\pi }{k_{\mathrm{1}}}\left[
\pi _{0}\left( \mathrm{L}\right) -\left\vert p^{\mathrm{L}}\right\vert %
\right] \right\}  \label{svf.12}
\end{equation}%
both for bosons and fermions.

Considering positive $p_{0}>0$, we find that $N_{n}^{\mathrm{cr}}$\ can be
approximated by forms (\ref{svf.14}); see details in Appendix \ref{App.1}.
In this case, the substantial value of $N_{n}^{\mathrm{cr}}$ are formed in
the range%
\begin{equation}
eE/k_{2}\geq \left\vert \pi _{0}\left( \mathrm{R}\right) \right\vert /k_{2}>K
\label{svf.15b}
\end{equation}%
and has a form
\begin{equation}
N_{n}^{\mathrm{cr}}\approx \exp \left\{ -\frac{2\pi }{k_{\mathrm{2}}}\left[
\left\vert \pi _{0}\left( \mathrm{R}\right) \right\vert -\left\vert p^{%
\mathrm{R}}\right\vert \right] \right\} .  \label{svf.15}
\end{equation}

Consequently, the quantity $N_{n}^{\mathrm{cr}}$ is almost constant over the
wide range of energies $p_{0}$ for any given $\lambda $ satisfying Eq.~(\ref%
{svf.3}). When $h_{\mathrm{1}},h_{\mathrm{2}}\rightarrow \infty $, one
obtains the result in a constant uniform electric field \cite%
{Nikishov1,Nikishov2}.

The analysis presented above\ reveals that the dominant contributions for
particle creation by a slowly varying field occurs in the ranges of large
kinetic energies, whose differential quantities have the asymptotic forms (%
\ref{svf.12}) for $p_{0}<0$ and (\ref{svf.15}) for $p_{0}>0$. Therefore, one
may represent the total number (\ref{dtq.3}) as%
\begin{align}
& N^{\mathrm{cr}}=V_{\bot }Tn^{\mathrm{cr}}\,,\ \ n^{\mathrm{cr}}=\frac{%
J_{(d)}}{\left( 2\pi \right) ^{d-1}}\int_{\sqrt{\lambda }<K_{\bot }}d\mathbf{%
p}_{\bot }I_{\mathbf{p}_{\bot }},\ \ I_{\mathbf{p}_{\bot }}=I_{\mathbf{p}%
_{\bot }}^{\left( \mathrm{1}\right) }+I_{\mathbf{p}_{\bot }}^{\left( \mathrm{%
2}\right) },  \notag \\
& I_{\mathbf{p}_{\bot }}^{\left( \mathrm{1}\right) }=\int_{-eE/k_{\mathrm{1}%
}+\pi _{\perp }}^{0}dp_{0}N_{n}^{\mathrm{cr}}\approx \int_{Kk_{\mathrm{1}%
}}^{eE/k_{\mathrm{1}}}d\pi _{0}\left( \mathrm{L}\right) \exp \left\{ -\frac{%
2\pi }{k_{\mathrm{1}}}\left[ \pi _{0}\left( \mathrm{L}\right) -\left\vert p^{%
\mathrm{L}}\right\vert \right] \right\} \,,  \notag \\
& I_{\mathbf{p}_{\bot }}^{\left( \mathrm{2}\right) }=\int_{0}^{eE/k_{\mathrm{%
2}}-\pi _{\perp }}dp_{0}N_{n}^{\mathrm{cr}}\approx \int_{Kk_{\mathrm{2}%
}}^{eE/k_{\mathrm{2}}}d\left\vert \pi _{0}\left( \mathrm{R}\right)
\right\vert \exp \left\{ -\frac{2\pi }{k_{\mathrm{2}}}\left[ \left\vert \pi
_{0}\left( \mathrm{R}\right) \right\vert -\left\vert p^{\mathrm{R}%
}\right\vert \right] \right\} \,.  \label{svf.16}
\end{align}%
Here $n^{\mathrm{cr}}$ presents the total number density of pairs created
per unit time and unit surface orthogonal to the electric field direction.

As it is shown in Appendix \ref{App.1}, the leading term reads

\begin{equation}
n^{\mathrm{cr}}=r^{\mathrm{cr}}\left( \frac{1}{k_{\mathrm{1}}}+\frac{1}{k_{%
\mathrm{2}}}\right) G\left( \frac{d}{2},\pi \frac{m^{2}}{eE}\right) ,\;\;r^{%
\mathrm{cr}}=\frac{J_{(d)}\left( eE\right) ^{d/2}}{(2\pi )^{d-1}}\exp
\left\{ -\pi \frac{m^{2}}{eE}\right\} ,  \label{svf.22}
\end{equation}%
where function $G$ is given by Eq. (\ref{svf.20}). The density $r^{\mathrm{cr%
}}$ is known in the theory of constant uniform electric field as the
pair-production rate (see the $d$ dimensional case in Refs.~\cite%
{L-field,GavG96a}). The density given by Eq.~(\ref{svf.22}) coincides with
the number density of pairs\ created per unit space volume, $N^{\mathrm{cr}%
}/V_{\left( d-1\right) }$, due to the uniform peak electric field given by a
time-dependent potential $A_{x}\left( t\right) $; see Ref.~\cite{AdoGavGit16}%
.

We see that the dominant contributions to the number density $n^{\mathrm{cr}%
} $, given by Eq.~(\ref{svf.22}),\ is proportional to the total energy of a
pair created and then the magnitude of the potential step, $\pi_{0}\left(
\mathrm{L}\right) +\left\vert \pi_{0}\left( \mathrm{R}\right) \right\vert =%
\mathbb{U}$. This magnitude is equal to a work done on a charged particle by
the electric field under consideration. The same behavior we see for the
number density $n^{\mathrm{cr}}$ of pairs created due to the small-gradient
potential steps of the other known exactly solvable cases: the Sauter field
\cite{x-case} and the $L$-constant electric field \cite{L-field} with the
step magnitudes $\mathbb{U}_{\mathrm{S}}=2eEL_{\mathrm{S}}$ and $\mathbb{U}_{%
\mathrm{L}}=eEL$, respectively. In these cases we have%
\begin{align}
n^{\mathrm{cr}} & =L_{\mathrm{S}}\delta r^{\mathrm{cr}}\text{\ for Sauter
field,}  \notag \\
n^{\mathrm{cr}} & =Lr^{\mathrm{cr}}\text{\ for }L\text{-constant field,}
\label{svf.22a}
\end{align}
where $L$ is the length of the applied constant field, $\delta=\sqrt{\pi}%
\Psi\left( \frac{1}{2},\frac{2-d}{2};\pi\frac{m^{2}}{eE}\right) $, and $%
\Psi\left( a,b;x\right) $ is the confluent hypergeometric function \cite%
{BatE53}. All three cases can be considered as regularizations for an
uniform electric field. This fact allows one to compare pair creation
effects in such fields. Thus, for a given magnitude of the electric field $E$
one can compare, for example, the pair creation effects in fields with equal
step magnitude, or one can determine such step magnitudes\emph{\ }for which
particle creation effects are the same. In the latter case, equating the
densities $n^{\mathrm{cr}}$ for the Sauter field and for the peak field to
the density $n^{\mathrm{cr}}$ for the $L$-constant field, we find an
effective length of the fields in both cases,%
\begin{align}
\;L_{\mathrm{eff}} & =L_{\mathrm{S}}\delta\text{\ for Sauter field,}  \notag
\\
L_{\mathrm{eff}} & =\left( k_{1}^{-1}+k_{2}^{-1}\right) G\left( \frac {d}{2}%
,\pi\frac{m^{2}}{eE}\right) \text{\ for peak field}.  \label{Leff}
\end{align}
By the definition $L_{\mathrm{eff}}=L$ for the $L$-constant field. One can
say that the Sauter and the peak electric fields with the same $L_{\mathrm{%
eff}}$ are equivalent to the $L$-constant field in pair production.

Using the above considerations and Eq. (\ref{p}) we perform the summation
(integration) in Eq. (\ref{Pv}) and obtain the vacuum-to-vacuum probability,%
\begin{align}
& P_{v}=\exp\left( -\mu N^{\mathrm{cr}}\right) ,\ \ \mu=\sum_{l=0}^{\infty }%
\frac{\kappa^{l}\epsilon_{l+1}}{(l+1)^{d/2}}\exp\left( -l\pi\frac{m^{2}}{eE}%
\right) \;,  \notag \\
& \epsilon_{l}=G(\frac{d}{2};\pi l\frac{m^{2}}{eE})\left[ G(\frac{d}{2};%
\frac{\pi m^{2}}{eE})\right] ^{-1},  \label{svf.24}
\end{align}
where $N^{\mathrm{cr}}=V_{\bot}Tn^{\mathrm{cr}}$ \ and $n^{\mathrm{cr}}$ is
given by Eq.~(\ref{svf.22}). Previously, similar results were obtained for
the Sauter field \cite{x-case} and the $L$-constant fields \cite{L-field}
with the corresponding $n^{\mathrm{cr}}$, given by Eq.~(\ref{svf.22a}), and
\begin{align}
\epsilon_{l} & =\epsilon_{l}^{\mathrm{L}}=1\;\mathrm{for\;L}\text{\textrm{-}}%
\mathrm{constant\ field},  \notag \\
\epsilon_{l} & =\epsilon_{l}^{\mathrm{S}}=\delta^{-1}\sqrt{\pi}\Psi\left(
\frac{1}{2},\frac{2-d}{2};l\pi\frac{m^{2}}{eE}\right) \;\mathrm{for\;Sauter\
field}.  \label{svf.25}
\end{align}

%%%%%%%%%%%%%%%%%%%%%%%%%%%%%%%%%%%%%%%%%%%%%%

\section{Very asymmetric peak\label{exp_dec_field}}

In the examples considered before \cite{x-case,L-field} and above,
increasing and decreasing parts of the electric field are near symmetric.
Here we consider an essentially asymmetric configuration of the step. We
suppose that the field grow from zero to its maximum value at the origin $%
x=0 $ very rapidly (that is, $k_{\mathrm{1}}$ is sufficiently large), while
the value of parameter $k_{\mathrm{2}}>0$ remains arbitrary and includes the
case of a smooth decay. We assume that the corresponding asymptotic
potential energy, $U_{\mathrm{L}}$, given by Eq. (\ref{f.9a}), define finite
magnitude of the potential step $\Delta U_{\mathrm{1}}=-U_{\mathrm{L}}$ for
increasing part of the field. Note that due to the invariance of the mean
numbers $N_{n}^{\mathrm{cr}}$ under the simultaneous change $k_{\mathrm{1}%
}\leftrightarrows k_{\mathrm{2}}$ and $\pi _{0}\left( \mathrm{L}\right)
\leftrightarrows -\pi _{0}\left( \mathrm{R}\right) $, one can easily
transform this situation to the case with a large $k_{\mathrm{2}}$ and
arbitrary $k_{\mathrm{1}}>0$. \bigskip Let us assume that a sufficiently
large $k_{\mathrm{1}}$ satisfies the following inequalities at given $\Delta
U_{\mathrm{1}}$ and $\pi _{0}\left( \mathrm{L}\right) =p_{0}+\Delta U_{%
\mathrm{1}}$:
\begin{equation}
\left\vert \pi _{0}\left( \mathrm{L}\right) \right\vert /k_{\mathrm{1}}\ll 1.
\label{edf.1}
\end{equation}%
Making use of condition Eq.~(\ref{edf.1}), we can approximately present $%
\left\vert \Delta \right\vert ^{2}$, given by Eq.~(\ref{de.21}), as
\begin{equation}
\left\vert \Delta \right\vert ^{2}\approx \left\vert \Delta _{\mathrm{ap}%
}\right\vert ^{2}=e^{-i\pi \nu _{2}}\left. \left\vert \left[ -\chi \Delta U_{%
\mathrm{1}}+\left\vert p^{\mathrm{L}}\right\vert -\left\vert p^{\mathrm{R}%
}\right\vert +k_{\mathrm{2}}h_{\mathrm{2}}\left( \frac{1}{2}+\frac{d}{d\eta
_{\mathrm{2}}}\right) \right] \Phi \left( 1-a_{\mathrm{2}},2-c_{\mathrm{2}%
};-\eta _{\mathrm{2}}\right) \right\vert ^{2}\right\vert _{x=0}\,.
\label{edf.2}
\end{equation}%
and finally obtain
\begin{equation}
\left\vert g\left( _{-}\left\vert ^{+}\right. \right) \right\vert
^{-2}\approx \left\{
\begin{array}{l}
\left\vert \mathcal{C}\Delta _{\mathrm{ap}}\right\vert ^{-2}\ \mathrm{for\
fermions} \\
\left\vert \mathcal{C}_{\mathrm{sc}}\left. \Delta _{\mathrm{ap}}\right\vert
_{\chi =0}\right\vert ^{-2}\ \mathrm{for\ bosons}%
\end{array}%
\right. \,  \label{5.9}
\end{equation}%
for the ranges $\Omega _{1}$, $\Omega _{3}$, and $\Omega _{5}$.

In an asymmetric case with $k_{\mathrm{1}}\gg k_{\mathrm{2}}$, we have $%
eE=k_{\mathrm{1}}\Delta U_{\mathrm{1}}$ at given $\Delta U_{\mathrm{1}}$
that implies $eE/k_{\mathrm{2}}\gg \Delta U_{\mathrm{1}}$. Then one can
disregard the term $\Delta U_{\mathrm{1}}$ in the leading-term-approximation
of $\left\vert g\left( _{-}\left\vert ^{+}\right. \right) \right\vert ^{-2}$
given by Eq.~(\ref{5.9}) for the ranges $\Omega _{1}$ and $\Omega _{5}$,
that is, outside of the Klein zone. Such an approximation does not depend on
the details of the field growth at $x<0$. We see that the relative
reflection $\left\vert R_{\zeta ,n}\right\vert ^{2}$ and transition $%
\left\vert T_{\zeta ,n}\right\vert ^{2}$ probabilities in the
leading-term-approximation are the same with ones produced due to the
exponentially decaying electric field, given by the potential (\ref{f.10}).

Let us consider the most asymmetric case when Eqs.~(\ref{5.9}) hold and the
parameter $k_{\mathrm{2}}$ is sufficiently small,%
\begin{equation}
h_{\mathrm{2}}\gg \max \left( 1,m^{2}/eE\right) \,.  \label{5.11}
\end{equation}%
In this case the exponentially decaying field (\ref{f.10}) is a
small-gradient field and we are interesting in the Klein zone, $\Omega _{3}$%
, where $N_{n}^{\mathrm{cr}}=\left\vert g\left( _{-}|^{+}\right) \right\vert
^{-2}$.

\bigskip Taking into account that both $\pi _{0}\left( \mathrm{L}\right) $
and $\pi _{0}\left( \mathrm{R}\right) $ satisfy the inequalities given by
Eq.~(\ref{Omega}) in the range $\Omega _{3}$, we see that $\pi _{0}\left(
\mathrm{R}\right) $ varies greatly,
\begin{equation}
\pi _{\perp }\leq -\pi _{0}\left( \mathrm{R}\right) \leq eE/k_{\mathrm{2}%
}+\Delta U_{\mathrm{1}}-\pi _{\perp }.  \label{5.13}
\end{equation}%
Note that in this range $2\pi _{\bot }<eE/k_{\mathrm{2}}+\Delta U_{\mathrm{1}%
}$. It can be seen from the asymptotic behavior of a confluent
hypergeometric function that $N_{n}^{\mathrm{cr}}$ is exponentially small if
both $\left\vert \pi _{0}\left( \mathrm{R}\right) \right\vert $ and $\pi
_{\perp }$ are large $\sim eE/k_{\mathrm{2}}$ with $\pi _{\perp }/\left\vert
\pi _{0}\left( \mathrm{R}\right) \right\vert \sim 1$ such that $\left\vert
p^{\mathrm{R}}\right\vert \ll \left\vert \pi _{0}\left( \mathrm{R}\right)
\right\vert $. Then the range of fixed $\pi _{\perp }$ is of interest in the
range (\ref{5.13}) and we assume that the inequality (\ref{svf.3}) holds, in
which $K_{\bot }$ is any given number satisfying the condition%
\begin{equation}
h_{\mathrm{2}}\gg K_{\bot }^{2}\gg \max \left( 1,m^{2}/eE\right) \,.
\label{5.12}
\end{equation}

Using asymptotic expressions of the confluent hypergeometric functions, we
find that the differential mean numbers of created pairs $N_{n}^{\mathrm{cr}%
} $, given by Eq.~(\ref{5.9}), can be approximated by distributions that\
vary from Eq.~(\ref{5.16}) to Eq.~(\ref{5.19a}); see details in Appendix \ref%
{App.1}. The only distribution (\ref{5.16}) depends on $\Delta U_{\mathrm{1}%
} $. However, the range of transverse momenta is quite tiny for this
distribution. It implies that in the leading term approximation{\large \ }%
the total number $N^{\mathrm{cr}}$\ of pairs created does not depend on%
{\large \ }$\Delta U_{\mathrm{1}}${\large \ }and therefore does not feel
peculiarities of the field growth at $x<0$\ in this approximation.

We see that $N_{n}^{\mathrm{cr}}$ given by Eq.~(\ref{5.19a}) are
exponentially small if $\left\vert \pi _{0}\left( \mathrm{R}\right)
\right\vert \sim \pi _{\perp }$. Then the substantial value of $N_{n}^{%
\mathrm{cr}}$ are formed in the range%
\begin{equation}
K_{\bot }\ll K<-2\pi _{0}\left( \mathrm{R}\right) /k_{2}<\left(
1-\varepsilon \right) h_{\mathrm{2}},  \label{D}
\end{equation}%
where \ $K$ is any given number and $K_{\bot }$ satisfies the inequalities (%
\ref{svf.3}) and (\ref{5.12}). In this range $\left\vert \pi _{0}\left(
\mathrm{R}\right) \right\vert \gg \pi _{\bot }$ and the distributions~(\ref%
{5.19a}) is approximated by Eq.~(\ref{5.17}) both for bosons and fermions.
Thus, Eq.~(\ref{5.17}) gives us the leading-term approximation for the
substantial value of $N_{n}^{\mathrm{cr}}$ over all the range (\ref{D}).
Note that the same distribution takes place in a small-gradient field for $%
p_{0}>0$, see Eq.~(\ref{svf.15}). Approximation (\ref{5.17}) does not depend
on the details of the field growth at $x<0$, therefore,{\large \ }it is the
same as{\large \ }in the case of the exponentially decaying electric field,
given by the potential (\ref{f.10}).

Using the above considerations, we can estimate dominant contributions to
the number density $n^{\mathrm{cr}}$\ of pairs created by the very
asymmetric peak as%
\begin{equation}
n^{\mathrm{cr}}=\frac{r^{\mathrm{cr}}}{k_{2}}G\left( \frac{d}{2},\pi \frac{%
m^{2}}{eE}\right) ,  \label{5.20}
\end{equation}%
where $r^{\mathrm{cr}}$ is given by Eq.~(\ref{svf.22}); see details in
Appendix \ref{App.1}. Thus, $n^{\mathrm{cr}}$ given by Eq.~(\ref{5.20}), is $%
k_{2}$-dependent part of the mean number density of pairs created in the
small-gradient field, given by Eq.~(\ref{svf.22}).

Finally, we can see that the vacuum-to-vacuum probability is%
\begin{equation}
P_{v}=\exp\left( -\mu N^{\mathrm{cr}}\right) ,  \label{5.21}
\end{equation}
where $N^{\mathrm{cr}}=V_{\bot}Tn^{\mathrm{cr}}$ \ and $n^{\mathrm{cr}}$ is
given by Eq.~(\ref{5.20}) and $\mu$ is given by Eq.~(\ref{svf.24}).

As it was mentioned above, the form of $N_{n}^{\mathrm{cr}}$ does not depend
on the details of the field growth at $x<0$ in the range of dominant
contribution.{\large \ }Therefore, calculations of total quantities in an
exponentially decaying field are quite representative for a large class of
the exponentially decaying electric fields switching on abruptly.

%%%%%%%%%%%%%%%%%%%%%%%%%%%%%%%%%%%%%%%%%%%%%

\section{Sharp peak\label{Sharp_peak}}

The choosing certain parameters of the peak field, one can obtain sharp
gradient fields that exist only in a small area in a vicinity of the origin $%
x=0$. The latter fields grows and/or decays rapidly\ near the point $x=0$.

Let us consider large parameters $k_{\mathrm{1}}$, $k_{\mathrm{2}%
}\rightarrow \infty $ with a fixed ratio $k_{\mathrm{1}}/k_{\mathrm{2}}$. We
assume that the corresponding asymptotic potential energies, $U_{\mathrm{R}}$
and $U_{\mathrm{L}}$, given by Eq. (\ref{f.9a}), define finite magnitudes of
the potential steps $\Delta U_{\mathrm{1}}$ and $\Delta U_{\mathrm{2}}$ for
increasing and decreasing parts,
\begin{equation}
-U_{\mathrm{L}}=\Delta U_{\mathrm{1}},\;\;U_{\mathrm{R}}=\Delta U_{2},
\label{spf.1}
\end{equation}%
respectively, and satisfy the following inequalities:
\begin{equation}
\Delta U_{\mathrm{1}}/k_{\mathrm{1}}\ll 1,\;\;\Delta U_{\mathrm{2}}/k_{%
\mathrm{2}}\ll 1.  \label{spf.2a}
\end{equation}%
In the ranges $\Omega _{1}$ and $\Omega _{5}$ the energy $\left\vert
p_{0}\right\vert $ is not restricted from the above, that is why in what
follows we consider only the subranges, where%
\begin{equation}
\;\;\max \left( \left\vert \pi _{0}\left( \mathrm{L}\right) \right\vert /k_{%
\mathrm{1}},\left\vert \pi _{0}\left( \mathrm{R}\right) \right\vert /k_{%
\mathrm{2}}\right) \ll 1.  \label{spf.2b}
\end{equation}%
In the range $\Omega _{3}$ for any given $\pi _{\bot }$ the absolute values
of $\left\vert p^{\mathrm{R}}\right\vert $ and $\left\vert p^{\mathrm{L}%
}\right\vert $ are restricted from above, see (\ref{g8}). Therefore,
condition (\ref{spf.2a}) implies Eq. (\ref{spf.2b}). This case corresponds
to a very sharp peak of the electric field with a given step magnitude $%
\mathbb{U=}\Delta U_{\mathrm{1}}+\Delta U_{\mathrm{2}}$. At the same time
this configuration imitates a sufficiently high rectangular potential step
[the Klein step; see Ref.~\cite{x-case} for details and the resolution of
the Klein paradox] and coincides with it as $k_{\mathrm{1}}$, $k_{\mathrm{2}%
}\rightarrow \infty $. Thus, this potential step can be considered as a
regularization of the Klein step. We have to compare this regularization
with another one presented by the Sauter potential in Ref.~\cite{x-case}. In
the case under consideration the confluent hypergeometric function can be
approximated by the first two terms in Eq. (\ref{de.11}), which are $\Phi
\left( a,c;\eta \right) $ with $c\approx 1$ and $\ \ a\approx \left( 1-\chi
\right) /2$. Then in the ranges $\Omega _{1}$, $\Omega _{3}$, and $\Omega
_{5}$, the coefficient $\left\vert g\left( _{-}|^{+}\right) \right\vert
^{-2} $, given by Eq.~(\ref{de.21}) for fermions, can be presented in the
leading-term approximation as
\begin{equation}
\left\vert g\left( _{-}|^{+}\right) \right\vert ^{-2}\approx \frac{%
4\left\vert p^{\mathrm{L}}\right\vert \left\vert p^{\mathrm{R}}\right\vert
\left\vert \pi _{0}\left( \mathrm{R}\right) -\left\vert p^{\mathrm{R}%
}\right\vert \right\vert }{\left( \mathbb{U}+\left\vert p^{\mathrm{R}%
}\right\vert -\left\vert p^{\mathrm{L}}\right\vert \right) ^{2}\left\vert
\pi _{0}\left( \mathrm{L}\right) +\left\vert p^{\mathrm{L}}\right\vert
\right\vert }.\   \label{spf.3}
\end{equation}%
This leading-term does not depend on $k_{\mathrm{1}}$ and $k_{\mathrm{2}}$.
Note that in the ranges $\Omega _{1}$ and $\Omega _{5}$, the coefficient $%
\left\vert g\left( _{-}|^{+}\right) \right\vert ^{-2}$ \ determinate the
relative reflection $\left\vert R_{\zeta ,n}\right\vert ^{2}$ and transition
$\left\vert T_{\zeta ,n}\right\vert ^{2}$ probabilities in the form (\ref%
{sc.2}) while in the range $\Omega _{3}$ it gives the differential mean
number of pairs created, $N_{n}^{\mathrm{cr}}=\left\vert g\left(
_{-}|^{+}\right) \right\vert ^{-2}$. Of course, $N_{n}^{\mathrm{cr}}\leq 1$,
while $\left\vert g\left( _{-}|^{+}\right) \right\vert ^{-2}$ is unbounded
in the ranges $\Omega _{1}$ and $\Omega _{5}$. For example, $\left\vert
g\left( _{-}|^{+}\right) \right\vert ^{-2}\approx \pi _{\bot }^{2}/\mathbb{U}%
^{2}$ if $\pi _{0}\left( \mathrm{R}\right) \gg \pi _{\bot }^{2}$ .

For bosons in the ranges $\Omega_{1}$, $\Omega_{3}$, and $\Omega_{5}$, the
leading-term approximation of $\left\vert g\left( _{-}|^{+}\right)
\right\vert ^{-2}$, given by Eq.~(\ref{de.25}), is%
\begin{align}
& \left\vert g\left( _{-}|^{+}\right) \right\vert ^{-2}\approx \frac{%
4\left\vert p^{\mathrm{L}}\right\vert \left\vert p^{\mathrm{R}}\right\vert }{%
\left( \left\vert p^{\mathrm{L}}\right\vert -\left\vert p^{\mathrm{R}%
}\right\vert \right) ^{2}+b^{2}},\;  \notag \\
& b=\frac{2\Delta U_{\mathrm{1}}}{k_{\mathrm{1}}}\left[ \pi_{0}\left(
\mathrm{L}\right) -\frac{\Delta U_{\mathrm{1}}}{4}\right] +\frac{2\Delta U_{%
\mathrm{2}}}{k_{\mathrm{2}}}\left[ -\pi_{0}\left( \mathrm{R}\right) -\frac{%
\Delta U_{\mathrm{2}}}{4}\right] .  \label{spf.4a}
\end{align}
Taking into account that $\left\vert \left\vert p^{\mathrm{L}}\right\vert
-\left\vert p^{\mathrm{R}}\right\vert \right\vert >\mathbb{U}$ in the ranges
$\Omega_{1}$ and $\Omega_{5}$, we obtain that
\begin{equation}
\left\vert g\left( _{-}|^{+}\right) \right\vert ^{-2}\approx\frac {%
4\left\vert p^{\mathrm{L}}\right\vert \left\vert p^{\mathrm{R}}\right\vert }{%
\left( \left\vert p^{\mathrm{L}}\right\vert -\left\vert p^{\mathrm{R}%
}\right\vert \right) ^{2}}.  \label{spf.4b}
\end{equation}
In the range $\Omega_{3}$ the difference $\left\vert \left\vert p^{\mathrm{L}%
}\right\vert -\left\vert p^{\mathrm{R}}\right\vert \right\vert $ are
restricted from above by Eq.~(\ref{g8}) and can tend to zero. That is why
the differential mean number of boson pairs created, $N_{n}^{\mathrm{cr}%
}=\left\vert g\left( _{-}|^{+}\right) \right\vert ^{-2}$ given by Eq.~(\ref%
{spf.4a}), can be large. It has a maximum, $N_{n}^{\mathrm{cr}}=4\left\vert
p^{\mathrm{L}}\right\vert \left\vert p^{\mathrm{R}}\right\vert
/b^{2}\rightarrow\infty$ at $\left\vert p^{\mathrm{L}}\right\vert
-\left\vert p^{\mathrm{R}}\right\vert =0$. This is an indication of a big
backreaction effect at $\left\vert p^{\mathrm{L}}\right\vert -\left\vert p^{%
\mathrm{R}}\right\vert \rightarrow0$. In contrast with the Fermi case the $%
k_{\mathrm{1}}$, $k_{\mathrm{2}}$ -dependent term $b^{2}$ in Eq.~(\ref%
{spf.4a}) can be neglected only in the range where
\begin{equation}
b^{2}\ll\left( \left\vert p^{\mathrm{L}}\right\vert -\left\vert p^{\mathrm{R}%
}\right\vert \right) ^{2}.  \label{spf.4c}
\end{equation}
Under the latter condition, one obtain%
\begin{equation}
N_{n}^{\mathrm{cr}}\approx\frac{4\left\vert p^{\mathrm{L}}\right\vert
\left\vert p^{\mathrm{R}}\right\vert }{\left( \left\vert p^{\mathrm{L}%
}\right\vert -\left\vert p^{\mathrm{R}}\right\vert \right) ^{2}}.
\label{spf.4d}
\end{equation}
Thus, we see that the concept of a sharp peak in the scalar QED is limited
by the condition $\min\left( \Delta U_{\mathrm{1}}/k_{\mathrm{1}},\Delta U_{%
\mathrm{2}}/k_{\mathrm{2}}\right) \gtrsim1$ for the fields under
consideration. We do not see similar problem in the spinor QED.

If $k_{1}=k_{2}$ (in this case $\Delta U_{2}=\Delta U_{1}=\mathbb{U}/2$), we
can compare the above results with the regularization of the Klein step by
the Sauter potential; see Ref. \cite{x-case}. We see that both
regularizations are in agreement for bosons under condition (\ref{spf.4c}).
Both regularizations are in agreement for fermions in the range $\Omega_{3}$
if $\left\vert \left\vert p^{\mathrm{L}}\right\vert -\left\vert p^{\mathrm{R}%
}\right\vert \right\vert \ll\mathbb{U}$. For fermions in the ranges $%
\Omega_{1}$ if $\left\vert p^{\mathrm{L}}\right\vert \ll\pi_{\bot}$ and $%
\Omega_{5}$ if $\left\vert p^{\mathrm{R}}\right\vert \ll\pi_{\bot}$ we have $%
\left\vert \left\vert p^{\mathrm{L}}\right\vert -\left\vert p^{\mathrm{R}%
}\right\vert \right\vert \gg\mathbb{U}$ and obtain from Eq.~(\ref{spf.3})
that%
\begin{equation}
\left\vert g\left( _{-}|^{+}\right) \right\vert ^{-2}\approx\frac {%
4\left\vert p^{\mathrm{L}}\right\vert \left\vert p^{\mathrm{R}}\right\vert
\left\vert \pi_{0}\left( \mathrm{R}\right) \right\vert }{\left( \left\vert
p^{\mathrm{R}}\right\vert -\left\vert p^{\mathrm{L}}\right\vert \right)
^{2}\left\vert \pi_{0}\left( \mathrm{L}\right) \right\vert }.  \label{spf.5}
\end{equation}
In the nonrelativistic subrange, $\left\vert \pi_{0}\left( \mathrm{R/L}%
\right) \right\vert \gg\mathbb{U}$, the leading-term in Eq.~(\ref{spf.5})
has a form given by Eq.~(\ref{spf.4b}), that is, it is the same for fermions
and bosons and both regularizations are in agreement. To compare our exact
results with results of the nonrelativistic consideration for a noncritical
rectangular step, $\mathbb{U<}2m$, (in this case the range $\Omega_{3}$ does
not exist) obtained in any textbook for one dimensional quantum motion
(e.g., see \cite{LanLiQM}), one set $p_{\bot}=0,$\ then $\pi_{\bot}=m,$ $\pi
_{0}\left( \mathrm{L}\right) =p_{0}=m+E$, and $\pi_{0}\left( \mathrm{R}%
\right) =p_{0}-\mathbb{U}=m+E-\mathbb{U}$.

%%%%%%%%%%%%%%%%%%%%%%%%%%%%%%%%%%%%%%%%%%%

\section{Concluding remarks}

We have presented new exactly solvable cases available in the nonperurbative
QED with $x$-electric potential steps that were formulated recently in Ref.
\cite{x-case}. In particular, we have considered in details three new
configurations of $x$-electric potential steps: a smooth peak, a strongly
asymmetric peak, and a sharp peak. Thus, together, with two recently
presented exactly solvable cases available in the QED with the steps (QED
with the Sauter field \cite{x-case} and with a constant electric field
between two capacitor plates \cite{L-field}), the most important physically
exactly solvable cases in such QED are described explicitly at present. We
note that varying parameters defining these steps it is possible to imitate,
at least qualitatively, a wide class of physically actual configurations of $%
x$-electric potential steps (constant electromagnetic inhomogeneous fields)
and calculate nonperturbatively various quantum vacuum effects in such
fields.

\subparagraph{\protect\large Acknowledgement}

The reported study of S.P.G., D.M.G., and A.A.Sh. was supported by a grant
from the Russian Science Foundation, Research Project No. 15-12-10009.

\appendix

\section{Mathematical details of study in the Klein zone \label{App.1}}

\subsection{Small-gradient field}

Considering negative energies $p_{0}\leq 0$, the total range for $\pi
_{0}\left( \mathrm{L}\right) $ given by Eq.~(\ref{svf.5}) can be divided in
four subranges
\begin{align}
\mathrm{(a)}& \;h_{\mathrm{1}}\geq 2\pi _{0}\left( \mathrm{L}\right) /k_{%
\mathrm{1}}>h_{\mathrm{1}}\left[ 1-\left( \sqrt{h_{\mathrm{1}}}g_{\mathrm{2}%
}\right) ^{-1}\right] ,\ \   \notag \\
\mathrm{(b)}& \;h_{\mathrm{1}}\left[ 1-\left( \sqrt{h_{\mathrm{1}}}g_{%
\mathrm{2}}\right) ^{-1}\right] \geq 2\pi _{0}\left( \mathrm{L}\right) /k_{%
\mathrm{1}}>h_{\mathrm{1}}\left( 1-\varepsilon \right) ,\ \   \notag \\
\mathrm{(c)}& \;h_{\mathrm{1}}\left( 1-\varepsilon \right) \geq 2\pi
_{0}\left( \mathrm{L}\right) /k_{\mathrm{1}}>h_{\mathrm{1}}/g_{\mathrm{1}},\
\   \notag \\
\mathrm{(d)}& \;h_{\mathrm{1}}/g_{\mathrm{1}}\geq 2\pi _{0}\left( \mathrm{L}%
\right) /k_{\mathrm{1}}\geq 2\pi _{\perp }/k_{\mathrm{1}},\ \   \label{svf.7}
\end{align}%
where $g_{\mathrm{1}}$, $g_{\mathrm{2}}$, and $\varepsilon $ are any given
numbers satisfying the conditions $g_{\mathrm{1}}\gg 1$, $g_{\mathrm{2}}$ $%
\gg 1$, and $\left( \sqrt{h_{\mathrm{1}}}g_{\mathrm{2}}\right) ^{-1}\ll
\varepsilon \ll 1$. Note that $\tau _{\mathrm{1}}=ih_{\mathrm{1}}/c_{\mathrm{%
1}}\approx h_{\mathrm{1}}k_{\mathrm{1}}/2\left\vert \pi _{0}\left( \mathrm{L}%
\right) \right\vert $ in the subranges \textrm{(a)}, \textrm{(b)}, and
\textrm{(c)}, and $\tau _{\mathrm{2}}=-ih_{\mathrm{2}}/\left( 2-c_{\mathrm{2}%
}\right) \approx h_{\mathrm{2}}k_{\mathrm{2}}/2\left\vert \pi _{0}\left(
\mathrm{R}\right) \right\vert $ in the whole range (\ref{svf.5}).

In these subranges we have for $\tau_{\mathrm{2}}^{-1}$ that
\begin{align}
& \mathrm{(a)}\ 1\leq\tau_{\mathrm{2}}^{-1}<1+\left( \sqrt{h_{\mathrm{2}}}g_{%
\mathrm{2}}\right) ^{-1},\   \notag \\
& \mathrm{(b)}\ \left[ 1+\left( \sqrt{h_{\mathrm{2}}}g_{\mathrm{2}}\right)
^{-1}\right] <\tau_{\mathrm{2}}^{-1}<1+\varepsilon k_{\mathrm{2}}/k_{\mathrm{%
1}},\   \notag \\
& \mathrm{(c)}\ 1+\varepsilon k_{\mathrm{2}}/k_{\mathrm{1}}<\tau_{\mathrm{2}%
}^{-1}<1+\left( 1-1/g_{\mathrm{1}}\right) k_{\mathrm{2}}/k_{\mathrm{1}},\
\notag \\
& \mathrm{(d)\ }1+\left( 1-1/g_{\mathrm{1}}\right) k_{\mathrm{2}}/k_{\mathrm{%
1}}<\tau_{\mathrm{2}}^{-1}\lesssim1+k_{\mathrm{2}}/k_{\mathrm{1}}.
\label{svf.8}
\end{align}
We see that $\tau_{\mathrm{1}}-1\rightarrow0$ and $\tau_{\mathrm{2}%
}-1\rightarrow0$ in the range \textrm{(a)}, while $\left\vert \tau _{\mathrm{%
1}}-1\right\vert \ $is some finite number in the range \textrm{(c)}, and $%
\left\vert \tau_{\mathrm{2}}-1\right\vert \ $some finite number in the
ranges \textrm{(c)} and \textrm{(d)}. In the range \textrm{(b)} these
quantities vary from their values in the ranges \textrm{(a)} and \textrm{(c)}%
.\

We choose $\chi =1$ for convenience in the Fermi case. In the range \textrm{%
(a)} we can use the asymptotic expression of the confluent hypergeometric
function given by Eq.~(\ref{eq.B1}) in the Appendix \ref{App.2}. Using Eqs.~(%
\ref{eq.B8}) and (\ref{eq.B9}), we finally find the leading term as%
\begin{equation}
N_{n}^{\mathrm{cr}}=e^{-\pi \lambda }\left[ 1+O\left( \left\vert \mathcal{Z}%
_{1}\right\vert \right) \right] ,  \label{svf.9}
\end{equation}%
for fermions and bosons, where $\max \left\vert \mathcal{Z}_{1}\right\vert
\lesssim g_{\mathrm{2}}^{-1}$. This expression in the leading order coincide
with the one for the case of uniform constant field \cite%
{Nikishov1,Nikishov2},%
\begin{equation}
N_{n}^{\mathrm{uni}}=e^{-\pi \lambda }.  \label{svf.uni}
\end{equation}%
In the range (c), the confluent hypergeometric function $\Phi \left(
1-a_{2},2-c_{2};-ih_{2}\right) $ is approximated by Eq.~(\ref{eq.B10a}) and
the function $\Phi \left( a_{1},c_{1};ih_{1}\right) $ is approximated by
Eq.~(\ref{eq.B10}) given in the Appendix \ref{App.2}. Then we find that%
\begin{equation}
N_{n}^{\mathrm{cr}}=e^{-\pi \lambda }\left[ 1+O\left( \left\vert \mathcal{Z}%
_{1}\right\vert \right) ^{-1}+O\left( \left\vert \mathcal{Z}_{2}\right\vert
\right) ^{-1}\right] ,  \label{svf.10}
\end{equation}%
where $\max \left\vert \mathcal{Z}_{1}\right\vert ^{-1}\lesssim \sqrt{g_{%
\mathrm{1}}/h_{\mathrm{1}}}$ and $\max \left\vert \mathcal{Z}_{2}\right\vert
^{-1}\lesssim g_{\mathrm{2}}^{-1}$. Using the asymptotic expression Eq.~(\ref%
{eq.B1}) and taking into account Eq.~(\ref{svf.9}) and (\ref{svf.10}), we
can estimate that $N_{n}^{\mathrm{cr}}\sim e^{-\pi \lambda }$ in the range
\textrm{(b)}.

In the range \textrm{(d)}, the confluent hypergeometric function $\Phi
\left( 1-a_{\mathrm{2}},2-c_{\mathrm{2}};-ih_{\mathrm{2}}\right) $ is
approximated by Eq.~(\ref{eq.B10a}) and the function $\Phi \left( a_{\mathrm{%
1}},c_{\mathrm{1}};ih_{\mathrm{1}}\right) $ is approximated by Eq.~(\ref%
{eq.B11}) given in the Appendix \ref{App.2}. In this range the differential
mean numbers in the leading-order approximation are%
\begin{align}
& N_{n}^{\mathrm{cr}}\approx \sinh \left( 2\pi \left\vert p^{\mathrm{L}%
}\right\vert /k_{\mathrm{1}}\right) \exp \left\{ -\frac{\pi }{k_{\mathrm{1}}}%
\left[ \pi _{0}\left( \mathrm{L}\right) -\left\vert p^{\mathrm{L}%
}\right\vert \right] \right\}  \notag \\
& \times \left\{
\begin{array}{l}
\sinh \left\{ \pi \left[ \pi _{0}\left( \mathrm{L}\right) +\left\vert p^{%
\mathrm{L}}\right\vert \right] /k_{\mathrm{1}}\right\} ^{-1}\ \mathrm{for\
fermions} \\
\cosh \left\{ \pi \left[ \pi _{0}\left( \mathrm{L}\right) +\left\vert p^{%
\mathrm{L}}\right\vert \right] /k_{\mathrm{1}}\right\} ^{-1}\ \mathrm{for\
bosons}%
\end{array}%
\right. \,.  \label{svf.11}
\end{align}%
It is clear that $N_{n}^{\mathrm{cr}}$ given by Eq.~(\ref{svf.11}) tends to
Eq.~(\ref{svf.10}), $N_{n}^{\mathrm{cr}}\rightarrow e^{-\pi \lambda }$, when
$\pi _{0}\left( \mathrm{L}\right) \gg \pi _{\bot }$. Consequently, the forms
(\ref{svf.11}) are valid in the whole range (\ref{svf.7}).

Considering positive $p_{0}>0$ and using the inequalities (\ref{Omega}) in
the range $\Omega _{3}$, we see that negative $\pi _{0}\left( \mathrm{R}%
\right) $ varies greatly while $\pi _{0}\left( \mathrm{L}\right) $ is
positive and very large,%
\begin{equation}
\pi _{\perp }\leq -\pi _{0}\left( \mathrm{R}\right) <\frac{eE}{k_{\mathrm{2}}%
},\;\frac{eE}{k_{\mathrm{1}}}<\pi _{0}\left( \mathrm{L}\right) \leq \frac{eE%
}{k_{\mathrm{2}}}+\frac{eE}{k_{\mathrm{1}}}-\pi _{\perp }.  \label{svf.13}
\end{equation}%
Taking into account that exact $N_{n}^{\mathrm{cr}}$ and its range of
formation is invariant under the simultaneous exchange $k_{\mathrm{1}%
}\leftrightarrows k_{\mathrm{2}}$ and $\pi _{0}\left( \mathrm{L}\right)
\leftrightarrows -\pi _{0}\left( \mathrm{R}\right) $, we find for $p_{0}>0$
that the differential mean numbers in the leading-order approximation are
\begin{eqnarray}
N_{n}^{\mathrm{cr}} &\approx &\sinh \left( 2\pi \left\vert p^{\mathrm{R}%
}\right\vert /k_{\mathrm{2}}\right) \exp \left\{ \frac{\pi }{k_{\mathrm{2}}}%
\left[ \pi _{0}\left( \mathrm{R}\right) +\left\vert p^{\mathrm{R}%
}\right\vert \right] \right\}  \notag \\
&&\times \left\{
\begin{array}{c}
\sinh \left\{ \pi \left[ \left\vert p^{\mathrm{R}}\right\vert -\pi
_{0}\left( \mathrm{R}\right) \right] /k_{\mathrm{2}}\right\} ^{-1}\ \mathrm{%
for\ fermions} \\
\cosh \left\{ \pi \left[ \left\vert p^{\mathrm{R}}\right\vert -\pi
_{0}\left( \mathrm{R}\right) \right] /k_{\mathrm{2}}\right\} ^{-1}\ \mathrm{%
for\ bosons}%
\end{array}%
\right. \,.  \label{svf.14}
\end{eqnarray}

Let us find the dominant contributions to the number density $n^{\mathrm{cr}%
} $ given by Eq.~(\ref{svf.16}). Using the variable changes,%
\begin{equation*}
s=\frac{2}{k_{\mathrm{1}}\lambda }\left[ \pi _{0}\left( \mathrm{L}\right)
-\left\vert p^{\mathrm{L}}\right\vert \right] \,\ \mathrm{in\ }I_{\mathbf{p}%
_{\bot }}^{\left( 1\right) },\,\ s=\frac{2}{k_{\mathrm{2}}\lambda }\left[
\left\vert \pi _{0}\left( \mathrm{R}\right) \right\vert -\left\vert p^{%
\mathrm{R}}\right\vert \right] \,\,\ \mathrm{in\ }I_{\mathbf{p}_{\bot
}}^{\left( 2\right) },
\end{equation*}%
we respectively represent the quantities $I_{\mathbf{p}_{\bot }}^{\left(
1\right) }$ and $I_{\mathbf{p}_{\bot }}^{\left( 2\right) }$ as%
\begin{equation}
I_{\mathbf{p}_{\bot }}^{\left( \mathrm{1}\right) }\approx
\int_{1}^{s_{1}^{\max }}\frac{ds}{s}\left\vert p^{\mathrm{L}}\right\vert
e^{-\pi \lambda s}\,,\,\ I_{\mathbf{p}_{\bot }}^{\left( \mathrm{2}\right)
}\approx \int_{1}^{s_{2}^{\max }}\frac{ds}{s}\left\vert p^{\mathrm{R}%
}\right\vert e^{-\pi \lambda s}\,,  \label{svf.17}
\end{equation}%
where%
\begin{equation*}
s_{j}^{\max }=\frac{eE}{k_{j}^{2}K},\;j=1,2;\;\;\left\vert p^{\mathrm{L/R}%
}\right\vert =\frac{eE}{sk_{\mathrm{1/2}}}-\frac{1}{4}k_{\mathrm{1/2}%
}\lambda s.
\end{equation*}

Assuming $\min \left[ \left( m/k_{1}\right) ^{2},\left( m/k_{2}\right) ^{2}%
\right] \gg K$ we see that the leading term in $I_{\mathbf{p}_{\bot }}$ (\ref%
{svf.16}) takes the following final form:%
\begin{equation}
I_{\mathbf{p}_{\bot }}\approx \left( \frac{eE}{k_{\mathrm{1}}}+\frac{eE}{k_{%
\mathrm{2}}}\right) \int_{1}^{\infty }\frac{ds}{s^{2}}e^{-\pi \lambda
s}=eE\left( \frac{1}{k_{\mathrm{1}}}+\frac{1}{k_{\mathrm{2}}}\right) e^{-\pi
\lambda }G\left( 1,\pi \lambda \right) ,  \label{svf.19}
\end{equation}%
where%
\begin{equation}
G\left( \alpha ,x\right) =\int_{1}^{\infty }\frac{ds}{s^{\alpha +1}}%
e^{-x\left( s-1\right) }=e^{x}x^{\alpha }\Gamma \left( -\alpha ,x\right) ,
\label{svf.20}
\end{equation}%
and $\Gamma \left( -\alpha ,x\right) $ is the incomplete gamma function.
Neglecting an exponentially small contribution, one can extend the
integration limit over $\mathbf{p}_{\bot }$ in Eq.~(\ref{svf.16}) from $%
\sqrt{\lambda }<K_{\bot }$ to $\sqrt{\lambda }<\infty .$ Then calculating
the Gaussian integral, we finally obtain the expression (\ref{svf.22}).

\subsection{Very asymmetric peak}

The total range (\ref{5.13}) can be divided into following subranges:%
\begin{align}
\mathrm{(a)}& \;\left( 1+\varepsilon \right) h_{2}\leq -2\pi _{0}\left(
\mathrm{R}\right) /k_{2}\leq h_{2}\left[ 1+\frac{2\left( \Delta U_{\mathrm{1}%
}-\pi _{\perp }\right) }{k_{\mathrm{2}}h_{2}}\right] ,  \notag \\
\mathrm{(b)}& \;h_{2}\left[ 1+\left( \sqrt{h_{2}}g_{2}\right) ^{-1}\right]
\leq -2\pi _{0}\left( \mathrm{R}\right) /k_{2}<\left( 1+\varepsilon \right)
h_{2},  \notag \\
\mathrm{(c)}& \;h_{2}\left[ 1-\left( \sqrt{h_{2}}g_{2}\right) ^{-1}\right]
\leq -2\pi _{0}\left( \mathrm{R}\right) /k_{2}<h_{2}\left[ 1+\left( \sqrt{%
h_{2}}g_{2}\right) ^{-1}\right] ,  \notag \\
\mathrm{(d)}& \;\left( 1-\varepsilon \right) h_{2}\leq -2\pi _{0}\left(
\mathrm{R}\right) /k_{2}<h_{2}\left[ 1-\left( \sqrt{h_{2}}g_{2}\right) ^{-1}%
\right] ,  \notag \\
\mathrm{(e)}& \;h_{2}/g_{1}<-2\pi _{0}\left( \mathrm{R}\right) /k_{2}<\left(
1-\varepsilon \right) h_{2},  \notag \\
\mathrm{(f)}& \;2\pi _{\mathrm{\perp }}/k_{\mathrm{2}}\leq -2\pi _{0}\left(
\mathrm{R}\right) /k_{2}<h_{2}/g_{1},  \label{5.14}
\end{align}%
where $g_{2},$\ $g_{1}$\ and $\varepsilon $\ are any given numbers
satisfying the conditions $g_{2}$\ $\gg 1,$\ $g_{\mathrm{1}}\gg 1$\ and $%
\varepsilon \ll 1$. We assume that $\varepsilon \sqrt{h_{2}}g_{2}\gg 1$.
There exists the range (a) if $\sqrt{2}\left( \Delta U_{\mathrm{1}}-\pi
_{\perp }\right) \gg \sqrt{eE}/g_{2}$\ and then we assume that $\varepsilon
<2\left( \Delta U_{\mathrm{1}}-\pi _{\perp }\right) /k_{\mathrm{2}}h_{2}$\
while the range (b) exists if $\sqrt{2}\left( \Delta U_{\mathrm{1}}-\pi
_{\perp }\right) >\sqrt{eE}/g_{2}$. If $\sqrt{2}\left( \Delta U_{\mathrm{1}%
}-\pi _{\perp }\right) <\sqrt{eE}/g_{2}$\ then there exists only the
subrange of the range (c) defined by an inequality $-\pi _{0}\left( \mathrm{R%
}\right) <eE/k_{\mathrm{2}}+\Delta U_{\mathrm{1}}-\pi _{\perp }$. Note that $%
\tau _{\mathrm{2}}=-ih_{\mathrm{2}}/(2-c_{\mathrm{2}})\approx h_{\mathrm{2}%
}k_{\mathrm{2}}/2\left\vert \pi _{0}\left( \mathrm{R}\right) \right\vert $\
in subranges from (a) to (e) and $\tau _{2}$\ varies from $1-O\left(
h_{2}^{-1}\right) $\ to $g_{1}$.

In the range (a) if it exists, the confluent hypergeometric function $\Phi
\left( 1-a_{\mathrm{2}},2-c_{\mathrm{2}};-\eta _{\mathrm{2}}\right) $\ is
approximated by Eq.~(\ref{eq.B10a}) given in Appendix \ref{App.2}. In this
range the differential mean numbers in the leading-order approximation are
very small,
\begin{equation}
N_{n}^{\mathrm{cr}}\approx \frac{2}{\left( h_{2}\right) ^{2}}\left[
1+O\left( \left\vert \mathcal{Z}_{2}\right\vert ^{-1}\right) \right] \
\times \left\{
\begin{array}{l}
\frac{\left\vert p^{\mathrm{L}}\right\vert }{\pi _{0}\left( \mathrm{L}%
\right) +\left\vert p^{\mathrm{L}}\right\vert }\ \mathrm{for\ fermions} \\
\frac{4\left\vert p^{\mathrm{L}}\right\vert }{k_{\mathrm{2}}}\ \mathrm{for\
bosons}%
\end{array}%
\right. \,,  \label{a}
\end{equation}%
where $\max \left\vert \mathcal{Z}_{2}\right\vert ^{-1}\sim \left(
\varepsilon \sqrt{h_{2}}\right) ^{-1}$.

In the range (c), $\tau _{2}-1\rightarrow 0$\ and, using Eqs.~ (\ref{eq.B5}%
), (\ref{eq.B2}) and (\ref{eq.B4}) from Appendix \ref{App.2} we find that%
\begin{equation}
N_{n}^{\mathrm{cr}}\approx \left\vert p^{\mathrm{L}}\right\vert \exp \left[ -%
\frac{\pi \lambda }{4}\right] \left[ 1+O\left( \left\vert \mathcal{Z}%
_{2}\right\vert ^{-1}\right) \right] \left\{
\begin{array}{c}
\frac{2}{\left\vert \left\vert p^{\mathrm{L}}\right\vert +\pi _{0}\left(
\mathrm{L}\right) \right\vert \cosh \left( \pi \lambda /4\right) }\ \
\mathrm{for\ fermions} \\
\frac{\left\vert \Gamma \left( \frac{1}{4}+\frac{i\lambda }{4}\right)
\right\vert ^{2}}{\sqrt{eE}\pi }\ \ \mathrm{for\ bosons}%
\end{array}%
\right. .  \label{5.16}
\end{equation}%
Note that $N_{n}^{\mathrm{cr}}$\ given by Eq.~(\ref{5.16}) are finite and
restricted, $N_{n}^{\mathrm{cr}}\leq 1$\ for fermions and $N_{n}^{\mathrm{cr}%
}\lesssim 1/g_{2}$\ for bosons. This form depends on $\Delta U_{\mathrm{1}}$%
. The range of transverse momenta is quite tiny here. In the range (b) if it
exists the distributions $N_{n}^{\mathrm{cr}}$\ vary between their values in
the ranges (a) and (c).

In the range (e) parameters $\eta _{2}$\ and $c_{2}$\ are large with $a_{2}$%
\ fixed and $\tau _{2}>1$\ with $\arg \left( 2-c_{2}\right) <0$. In this
case, using the asymptotic expression of the confluent hypergeometric
function given by Eq.~(\ref{eq.B10}) in Appendix, we find that%
\begin{equation}
N_{n}^{\mathrm{cr}}=\exp \left\{ \frac{2\pi }{k_{\mathrm{2}}}\left[
\left\vert p^{\mathrm{R}}\right\vert +\pi _{0}\left( \mathrm{R}\right) %
\right] \right\} \left[ 1+O\left( \left\vert \mathcal{Z}_{2}\right\vert
^{-1}\right) \right] .  \label{5.17}
\end{equation}%
both for fermions and bosons, where $Z_{2}$\ is given by Eq.~(\ref{eq.B5}).
We note that\ modulus $\left\vert \mathcal{Z}_{2}\right\vert ^{-1}$\ varies
from $\left\vert \mathcal{Z}_{2}\right\vert ^{-1}\sim \left( \varepsilon
\sqrt{h_{2}}\right) ^{-1}$\ to $\left\vert \mathcal{Z}_{2}\right\vert
^{-1}\sim \left[ \left( g_{1}-1\right) \sqrt{h_{2}}\right] ^{-1}$.
Approximately, expression (\ref{5.17}) can be written as%
\begin{equation}
N_{n}^{\mathrm{cr}}\approx \exp \left( -\frac{\pi \pi _{\perp }^{2}}{k_{%
\mathrm{2}}\left\vert \pi _{0}\left( \mathrm{R}\right) \right\vert }\right) .
\label{5.18}
\end{equation}%
Note that $eE/g_{1}<k_{2}\left\vert \pi _{0}\left( \mathrm{R}\right)
\right\vert <\left( 1-\varepsilon \right) eE$\ in the range (e). It is clear
that the distribution $N_{n}^{\mathrm{cr}}$\ given by Eq.~(\ref{5.18}) has
the following limiting form:%
\begin{equation*}
N_{n}^{\mathrm{cr}}\rightarrow e^{-\pi \lambda }\ \ \mathrm{as\ \ }k_{%
\mathrm{2}}\left\vert \pi _{0}\left( \mathrm{R}\right) \right\vert
\rightarrow \left( 1-\varepsilon \right) eE\ .
\end{equation*}%
Thus, we see that the result in a constant uniform electric field, given by
Eq.~(\ref{svf.uni}) is reproduced in the wide range of high energies, $%
\left\vert \pi _{0}\left( \mathrm{R}\right) \right\vert \sim eE/k_{2}$. In
the range (d), the distributions $N_{n}^{\mathrm{cr}}$\ vary from their
values in the ranges (c) and (e) for fermions and bosons.

In the range (f), we can use the asymptotic expression of the confluent
hypergeometric function for large $h_{2}$\ with fixed $a_{2}$\ and $c_{2}$\
given by Eq.~(\ref{eq.B11}) in Appendix \ref{App.2} to show that the number
of particles created is%
\begin{align}
& N_{n}^{\mathrm{cr}}\approx \sinh \left( 2\pi \left\vert p^{\mathrm{R}%
}\right\vert /k_{\mathrm{2}}\right) \exp \left\{ \frac{\pi }{k_{\mathrm{2}}}%
\left[ \pi _{0}\left( \mathrm{R}\right) +\left\vert p^{\mathrm{R}%
}\right\vert \right] \right\}  \notag \\
& \times \left\{
\begin{array}{c}
\sinh \left\{ \pi \left[ \left\vert p^{\mathrm{R}}\right\vert -\pi
_{0}\left( \mathrm{R}\right) \right] /k_{\mathrm{2}}\right\} ^{-1}\ \mathrm{%
for\ fermions} \\
\cosh \left\{ \pi \left[ \left\vert p^{\mathrm{R}}\right\vert -\pi
_{0}\left( \mathrm{R}\right) \right] /k_{\mathrm{2}}\right\} ^{-1}\ \mathrm{%
for\ bosons}%
\end{array}%
\right. \,.  \label{5.19a}
\end{align}%
The same distribution takes place in a small-gradient field for $p_{0}>0$,
see Eq.~(\ref{svf.14}).

Let us find the dominant contributions to the total number $N^{\mathrm{cr}}$
of pairs created. To this end, we represent the leading terms of integral (%
\ref{dtq.3}) as a sum of two contributions, one due to the ranges (e) and%
\textrm{\ }(f) and another one due to the ranges (a),\ (b), (c), and (d):
\begin{align}
& N^{\mathrm{cr}}=V_{\left( d-1\right) }n^{\mathrm{cr}}\,,\ \ n^{\mathrm{cr}%
}=\frac{V_{\bot }\ TJ_{(d)}}{\left( 2\pi \right) ^{d-1}}\int_{\sqrt{\lambda }%
<K_{\bot }}d\mathbf{p}_{\bot }\ I_{\mathbf{p}_{\bot }},\ \ I_{\mathbf{p}%
_{\bot }}=I_{\mathbf{p}_{\bot }}^{\left( 1\right) }+I_{\mathbf{p}_{\bot
}}^{\left( 2\right) },  \notag \\
& I_{\mathbf{p}_{\bot }}^{\left( 1\right) }=\int_{\pi _{0}\left( \mathrm{R}%
\right) \in \mathrm{(a)\cup (b)\cup (c)\cup (d)}}d\pi _{0}\left( \mathrm{R}%
\right) N_{n}^{\mathrm{cr}}\,,\ \ I_{\mathbf{p}_{\bot }}^{\left( 2\right)
}=\int_{\pi _{0}\left( \mathrm{R}\right) \in \mathrm{(e)\cup (f)}}d\pi
_{0}\left( \mathrm{R}\right) N_{n}^{\mathrm{cr}}\,.  \label{edf.25}
\end{align}

The main contribution to the integral (\ref{edf.25}) is due to the wide
range (e)$\mathrm{\cup }$(f) of a large kinetic energy $\left\vert \pi
_{0}\left( \mathrm{R}\right) \right\vert $ with a relatively small
transverse momentum $\left\vert \mathbf{p}_{\perp }\right\vert $. The
contribution to this quantity from the relatively narrow momentum ranges \
(a),\ (b), (c), and (d)\ is finite and the corresponding integral $I_{%
\mathbf{p}_{\bot }}^{\left( 1\right) }$ is of the order $\sqrt{eE}/g_{2}$ .
The integral $I_{\mathbf{p}_{\bot }}^{\left( 2\right) }$ can be taken from
Eq.~(\ref{svf.16}) and approximated by the form from Eq.~(\ref{svf.17}).
Thus, the dominant contribution is given by integral (\ref{svf.19}) at $k_{%
\mathrm{1}}\rightarrow \infty $. Then calculating the Gaussian integral, we
find the form (\ref{5.20}).

\section{Some asymptotic expansions \label{App.2}}

The asymptotic expression of the confluent hypergeometric function for large
$\eta$ and $c$ with fixed $a$ and $\tau=\eta/c\sim1$ is given by
Eq.~(13.8.4) in \cite{DLMF} as%
\begin{align}
& \Phi\left( a,c;\eta\right) \simeq c^{a/2}e^{\mathcal{Z}^{2}/4}F\left(
a,c;\tau\right) ,\ \ \mathcal{Z=-}\left( \tau-1\right) \mathcal{W}\left(
\tau\right) \sqrt{c},  \notag \\
& F\left( a,c;\tau\right) =\tau\mathcal{W}^{1-a}D_{-a}\left( \mathcal{Z}%
\right) +\mathcal{R}D_{1-a}\left( \mathcal{Z}\right) ,  \notag \\
& \mathcal{R}=\left( \mathcal{W}^{a}-\tau\mathcal{W}^{1-a}\right) /\mathcal{Z%
},\ \ \mathcal{W}\left( \tau\right) =\left[ 2\left( \tau -1-\ln\tau\right)
/\left( \tau-1\right) ^{2}\right] ^{1/2}  \label{eq.B1}
\end{align}
where $D_{-a}\left( \mathcal{Z}\right) $ is the Weber parabolic cylinder
function (WPCF) \cite{BatE53}. Using Eq.~(\ref{eq.B1}) we present the
functions $y_{2}^{2}$, $y_{1}^{1}$ and their derivatives at $x=0$ as%
\begin{align}
& \left. y_{1}^{1}\right\vert _{x=0}\simeq e^{-ih_{1}/2}\left( ih_{1}\right)
^{\nu_{1}}\ c_{1}^{a_{1}/2}e^{\mathcal{Z}_{1}^{2}/4}F\left(
a_{1},c_{1};\tau_{1}\right) ,\   \notag \\
& \mathcal{Z}_{1}\mathcal{=}\mathcal{-}\left( \tau_{1}-1\right) \mathcal{W}%
\left( \tau_{1}\right) \sqrt{c_{1}},\ \ \tau_{1}=ih_{1}/c_{1},\   \notag \\
& \left. \frac{dy_{1}^{1}}{d\eta_{1}}\right\vert _{x=0}\simeq
e^{-ih_{1}/2}\left( ih_{1}\right) ^{\nu_{1}}c_{1}^{a_{1}/2}e^{\mathcal{Z}%
_{1}^{2}/4}\left[ -\frac{1}{2ih_{1}}+\frac{1}{c_{1}}\frac{\partial}{%
\partial\tau_{1}}\right] F\left( a_{1},c_{1};\tau_{1}\right) ;\   \notag \\
& \left. y_{2}^{2}\right\vert _{x=0}\simeq e^{ih_{2}/2}\left( ih_{2}\right)
^{-\nu_{2}}\left( 2-c_{2}\right) ^{\left( 1-a_{2}\right) /2}e^{\mathcal{Z}%
_{2}^{2}/4}F\left( 1-a_{2},2-c_{2};\tau_{2}\right) ,\   \notag \\
& \mathcal{Z}_{2}\mathcal{=}\mathcal{-}\left( \tau_{2}-1\right) \mathcal{W}%
\left( \tau_{2}\right) \sqrt{2-c_{2}},\ \ \tau_{2}=-ih_{2}/\left(
2-c_{2}\right) ,  \notag \\
& \left. \frac{dy_{2}^{2}}{d\eta_{2}}\right\vert _{x=0}\simeq
e^{ih_{2}/2}\left( ih_{2}\right) ^{-\nu_{2}}\left( 2-c_{2}\right) ^{\left(
1-a_{2}\right) /2}e^{\mathcal{Z}_{2}^{2}/4}\left[ -\frac{1}{2ih_{2}}-\frac{1%
}{2-c_{2}}\frac{\partial}{\partial\tau_{2}}\right] F\left(
1-a_{2},2-c_{2};\tau_{2}\right) .\   \label{eq.B5}
\end{align}

Assuming $\tau-1\rightarrow0$, one has%
\begin{align*}
& \mathcal{W}^{1-a}\approx1+\frac{a-1}{3}\left( \tau-1\right) ,\ \ \mathcal{R%
}\approx\frac{2\left( a+1\right) }{3\sqrt{c}},\ \ \mathcal{Z\approx-}\left(
\tau-1\right) \sqrt{c}, \\
& \frac{\partial F\left( a,c;\tau\right) }{\partial\tau}\approx\frac {2+a}{3}%
D_{-a}\left( \mathcal{Z}\right) +\frac{\partial D_{-a}\left( \mathcal{Z}%
\right) }{\partial\tau}+\mathcal{R}\frac{\partial D_{1-a}\left( \mathcal{Z}%
\right) }{\partial\tau}.
\end{align*}
Expanding WPCFs near $\mathcal{Z}=0$, in the leading approximation at $%
\mathcal{Z}\rightarrow0$ one obtains that%
\begin{align}
& \frac{\partial F\left( a,c;\tau\right) }{\partial\tau}\approx-\sqrt{\eta }%
D_{-a}^{\prime}\left( 0\right) +O\left( \eta\right) ,  \notag \\
& F\left( a,c;\tau\right) \approx D_{-a}\left( 0\right) +O\left(
c^{-1/2}\right) ,  \label{eq.B2}
\end{align}
and%
\begin{equation}
D_{-a}\left( 0\right) =\frac{2^{-a/2}\sqrt{\pi}}{\Gamma\left( \frac{a+1}{2}%
\right) },\ \ D_{-a}^{\prime}\left( 0\right) =\frac{2^{\left( 1-a\right) /2}%
\sqrt{\pi}}{\Gamma\left( \frac{a}{2}\right) },  \label{eq.B4}
\end{equation}
where $\Gamma(z)$ is the Euler gamma function. We find under condition (\ref%
{svf.1}) that
\begin{align}
& \left\vert p^{\mathrm{L}/\mathrm{R}}\right\vert \approx\left\vert \pi
_{0}\left( \mathrm{L/R}\right) \right\vert \left( 1-\lambda/h_{\mathrm{1}/%
\mathrm{2}}\right) ,\ \ a_{\mathrm{1}}\approx a_{\mathrm{2}}\approx\left(
1-\chi\right) /2-i\lambda/2,\ \   \notag \\
& c_{\mathrm{1}}\approx1+i\left( \frac{2\pi_{0}\left( \mathrm{L}\right) }{k_{%
\mathrm{1}}}-\lambda\right) ,\ \ 2-c_{\mathrm{2}}\approx1-i\left( \frac{%
2\left\vert \pi_{0}\left( \mathrm{R}\right) \right\vert }{k_{\mathrm{2}}}%
-\lambda\right) ,\ \   \notag \\
& \tau_{\mathrm{1}}-1\approx-\frac{1}{h_{\mathrm{1}}}\left( -i-\lambda +%
\frac{2p_{0}}{k_{\mathrm{1}}}\right) ,\ \ \ \tau_{\mathrm{2}}-1\approx \frac{%
1}{h_{\mathrm{2}}}\left( -i+\lambda+\frac{2p_{0}}{k_{\mathrm{2}}}\right) .\
\   \label{eq.B6}
\end{align}
Using Eqs.~(\ref{eq.B5}), (\ref{eq.B2}), and (\ref{eq.B6}) we represent Eq.~(%
\ref{de.A12}) in the form%
\begin{align}
N_{n}^{\mathrm{cr}} & =e^{-\pi\lambda/2}\left[ \left\vert \delta
_{0}\right\vert ^{-2}+O\left( h_{\mathrm{1}}^{-1/2}\right) +O\left( h_{%
\mathrm{2}}^{-1/2}\right) \right] ,  \notag \\
\delta_{0} & =e^{i\pi/4}D_{-a_{\mathrm{1}}}\left( 0\right) D_{a_{\mathrm{2}%
}-1}^{\prime}\left( 0\right) -e^{-i\pi/4}D_{-a_{\mathrm{1}}}^{\prime}\left(
0\right) D_{a_{\mathrm{2}}-1}\left( 0\right) .  \label{eq.B8}
\end{align}
Assuming $\chi=1$ for fermions and $\chi=0$ for bosons, and using the
relations of the Euler gamma function we find that
\begin{equation}
\delta_{0}=\exp\left( i\frac{3\pi}{4}-i\frac{\pi\chi}{2}\right)
e^{\pi\lambda/4}.  \label{eq.B9}
\end{equation}

Assuming $\left\vert \tau-1\right\vert \sim1$, one can use the asymptotic
expansions of WPCFs in Eq.~(\ref{eq.B1}), e.g., see \cite{BatE53,DLMF}. Note
that $\arg\left( \mathcal{Z}\right) \approx\frac{1}{2}\arg\left( c\right) $
if $1-\tau>0$. Then one finds that%
\begin{equation}
\Phi\left( a,c;\eta\right) =\left( 1-\tau\right) ^{-a}\left[ 1+O\left(
\left\vert \mathcal{Z}\right\vert ^{-1}\right) \right] \ \ \mathrm{if}\
\;1-\tau>0.  \label{eq.B10a}
\end{equation}
In the case of $1-\tau<0$, one has
\begin{equation*}
\arg\left( \mathcal{Z}\right) \approx\left\{
\begin{array}{c}
\frac{1}{2}\arg\left( c\right) +\pi\ \ \mathrm{if}\ \ \arg\left( c\right) <0
\\
\frac{1}{2}\arg\left( c\right) -\pi\ \ \mathrm{if}\ \ \arg\left( c\right) >0%
\end{array}
\right. .
\end{equation*}
Then one obtains finally that%
\begin{equation}
\Phi\left( a,c;\eta\right) =\left\{
\begin{array}{l}
\left( \tau-1\right) ^{-a}e^{-i\pi a}\left[ 1+O\left( \left\vert \mathcal{Z}%
\right\vert ^{-1}\right) \right] \ \ \mathrm{if}\ \;\arg\left( c\right) <0
\\
\left( \tau-1\right) ^{-a}e^{i\pi a}\left[ 1+O\left( \left\vert \mathcal{Z}%
\right\vert ^{-1}\right) \right] \ \ \mathrm{if}\ \;\arg\left( c\right) >0%
\end{array}
\right. .  \label{eq.B10}
\end{equation}

The asymptotic expression of the confluent hypergeometric function $%
\Phi\left( a,c;\pm ih\right) $ for large real $h$ with fixed $a$ and $c$ is
given by Eq.~(6.13.1(2)) in \cite{BatE53} as%
\begin{equation}
\Phi\left( a,c;\pm ih\right) =\frac{\Gamma\left( c\right) }{\Gamma\left(
c-a\right) }e^{\pm i\pi a/2}h^{-a}+\frac{\Gamma\left( c\right) }{%
\Gamma\left( a\right) }e^{\pm ih}\left( e^{\pm i\pi/2}h\right)
^{a-c}+O\left( \left\vert h\right\vert ^{-a-1}\right) +O\left( \left\vert
h\right\vert ^{a-c-1}\right) .  \label{eq.B11}
\end{equation}

\end{document}